\documentclass[journal]{IEEEtran}
\ifCLASSINFOpdf
\else
\fi
\usepackage{graphicx}
\usepackage{bm}
\usepackage{amsmath}
\usepackage{flushend}
\usepackage{cite}
\DeclareMathOperator{\sinc}{sinc}
\DeclareMathOperator{\Ei}{Ei}

\begin{document}

%
\title{Radiation from transmission lines PART I: free space transmission lines}
%
%
%


\author{
    \IEEEauthorblockN{Reuven Ianconescu\IEEEauthorrefmark{1}, Vladimir Vulfin\IEEEauthorrefmark{2}}\\
    \IEEEauthorblockA{\IEEEauthorrefmark{1}Shenkar College of Engineering and Design, Ramat Gan, Israel, riancon@gmail.com}\\
    \IEEEauthorblockA{\IEEEauthorrefmark{2}Ben-Gurion University of the Negev, Beer Sheva 84105, Israel, vlad2042@yahoo.com}
}

\maketitle


\begin{abstract}
This work derives exact expressions for the radiation from two conductors non isolated
TEM transmission lines of any cross section in free space. We cover the cases of infinite, semi-infinite and
finite transmission lines and show that while an infinite transmission line does not radiate,
there is a smooth transition between the radiation from a finite to a semi-infinite transmission
line. Our analysis is in the frequency domain and we consider transmission lines carrying
any combination of forward and backward waves. The analytic results are validated by successful
comparison with ANSYS commercial software simulation results, and successful comparisons with
other published results.
\end{abstract}


%
\IEEEpeerreviewmaketitle


\section{Introduction}

\begin{figure}[!tbh]
\includegraphics[width=9cm]{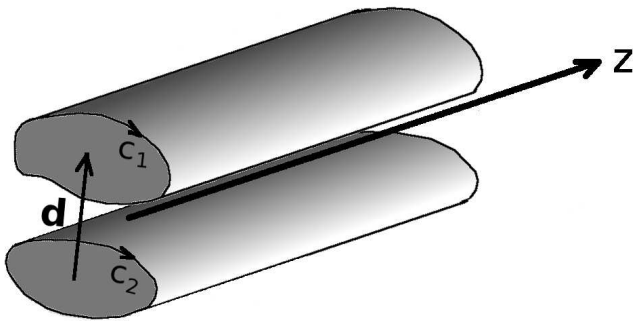}
\caption{A basic configuration of a two ideal conductors TL, with a well defined separation between the
conductors. The surface current distributions on the contours of the conductors is known from
electrostatic considerations, and given that for a two-conductors TL there is only one (differential) TEM mode,
the total current is the same on both conductors but with
opposite signs. The arrow shows the vector distance between the center of the surface current
distributions, named $\mathbf{d}$, obtained for a twin lead equivalent (see Appendix~A). $c_{1,2}$
are the contours of the ``upper'' and ``lower'' conductors, respectively. We consider the case of small
electric cross section $kd\ll 1$, $k$ is the wavenumber.}
\label{config}
\end{figure}

The aim of this work is to calculate the radiated power from two conductors transmission
lines (TL) in free space. We consider any cross section of small electric size (shown in Figure~\ref{config})
and TL of any length, analyzing the infinite, semi-infinite and finite TL cases. Part~II is the generalization
of this work for TL inside dielectric insulator and will be published separately.
Some preliminary results of this work have been presented in \cite{eumw_2016, icsee_2016}.

One of the earliest analysis of radiation from transmission lines (TL) is presented in the paper
``Radiation from Transmission Lines'' \cite{Manneback}, published in 1923.
In this publication the radiation from open ended twin lead TL in free space, at the resonance
frequencies is calculated.

A more conclusive and full analysis is presented in the 1951 paper
``Radiation Resistance of a Two-Wire Line'' \cite{storer}. This work calculates the radiation
resistance of a twin lead loaded at its termination by any impedance, considering TL ohmic losses as well.
In this work we consider 0 ohmic losses, but generalize \cite{storer} as follows:
\begin{itemize}
\item We consider the separate radiation of the forward and backward waves and also their combination. From such
analysis it comes out that the interference term between the waves does not contribute to the radiated power.

\item We analyze the radiation from semi-infinite TL, and show the connection with the finite TL case.

\item We do not limit ourselves to the twin lead cross section, and present a general algorithm for TL of any cross section.

\item We develop a more accurate radiation resistance.
\end{itemize}

Additional works on the subject can be found in \cite{Bingeman_2001,Nakamura_2006,carson}. It is
interesting to remark that \cite{msu} claims that balanced TL do not radiate. Although \cite{msu} is only an
educational document of a university, such inaccuracy is a symptom showing that the subject of radiation from TL
is not well enough known in the electromagnetic community, requiring additional research on this subject.

\begin{figure}[!tbh]
\includegraphics[width=9cm]{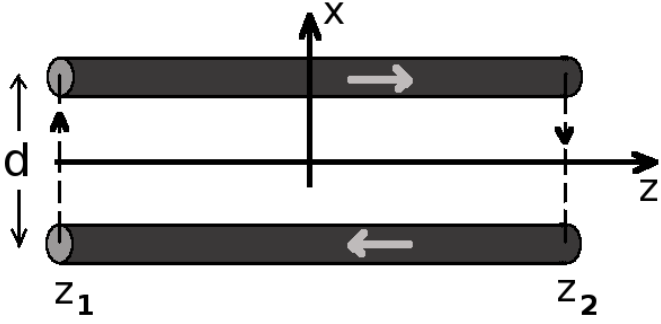}
\caption{The transmission line is modeled as a twin lead in free space, with distance $d$ between the
conductors. The currents in the transmission line flow in the $z$ direction at $x=\pm d/2$ and they
contribute to the magnetic vector potential $A_z$. The termination
currents (source or load) flow in the $x$ direction and contribute the
magnetic vector potential $A_x$. The arrows on the conductors show the conventional directions
of those currents. The wires appear in the figure with finite thickness, but are considered of
0 radius. The transmission line goes in the $z$ direction from $z_1$ to $z_2$.}
\label{conf_tl_rad}
\end{figure}

There are two appendices in this work. In Appendix~A we calculate the far potential vector from a general
cross section TL (see Figure~\ref{config}), and show that this far potential vector can be represented
in terms of an equivalent twin lead as shown in Figure~\ref{conf_tl_rad}. By ``far'' we mean in the transverse $x,y$
direction because this appendix is not limited to finite TL, so for being able to use the results for
a semi-infinite TL, we keep everything in cylindrical coordinates, and consider the TL between $z_1$
and $z_2$ on the $z$ axis. The twin lead model is defined without loss of generality on the $x,z$
plane and allows us to define $x$ directed termination currents as current filaments \cite{storer} across
the termination (see Figure~\ref{conf_tl_rad}).
Those termination currents contribute a far $x$ directed potential vector, calculated in Appendix~A.

In Appendix~B we show how to calculate for any cross section the parameters needed to determine the radiation:
the separation distance $d$ in the twin lead representation, and the characteristic impedance $Z_0$. We
perform this analysis on two cross section examples.

It is important to remark that the calculations in this work and in
\cite{storer,Manneback,Bingeman_2001,Nakamura_2006,carson} are completely different from
what is presented as ``traveling wave antenna'' in many antenna and electromagnetic books like
\cite{collin,orfanidis,ramo,jordan,balanis}. The last ones consider only the current in the
``upper'' conductor in Figure~\ref{conf_tl_rad}, while we consider {\it all 4} currents appearing
in the figure. This is not an attempt to criticize those works, but only to mention their results
do not represent radiation from transmission lines, hence are not comparable with the results
of this work or \cite{storer,Manneback,Bingeman_2001,Nakamura_2006,carson}.

It should be also mentioned that power loss from TL is also affected by nearby objects
interfering with the fields, line bends, irregularities, etc. This is certainly true, but those affect {\it not
only} the radiation, {\it but also} the basic, ``ideal'' TL model in what concerns the characteristic
impedance, the propagation wave number, etc. Those non-ideal phenomena are {\it not considered} in the current
work, nor in \cite{storer,Manneback,Bingeman_2001,Nakamura_2006,carson}, and also not in
\cite{collin,orfanidis,ramo,jordan,balanis}.

The methodology we use for calculating radiation losses is first order perturbation: we use the lossless
(0'th order solution) for the electric current to derive the losses, and we therefore use in Appendix~A
the $e^{-jkz}$ dependence. This methodology is used
to derive the ohmic and dielectric losses \cite{pozar,orfanidis,ramo,jordan}, and the same approach is used in 
different radiation schemes from free electrons: one uses the 0'th order current (which is
unaffected by the radiation) to calculate the radiation \cite{WEP085, MOP078, nima}. To be
mentioned that the same approach has been used in \cite{storer,Manneback,Bingeman_2001,Nakamura_2006,carson}
(although \cite{storer} discussed about higher order terms, without applying them).

The main text is organized as follows. In Section~II we use the results of Appendix~A to calculate
the power radiated from a finite TL carrying a forward wave current, and generalize this result for
any combination of waves. As mentioned earlier, the results of Appendix~A are applicable also for
semi-infinite TL, but we prefer to start with the finite TL, because in the following Section~\ref{validation} we
validate the analytic results of Section~\ref{finite} by comparing them
with ANSYS commercial software simulation results, and with published results obtained by other authors.

Section~\ref{inf_and_semiinf} we base on Appendix~A to analyze an infinite and semi-infinite TL. As
expected, an infinite TL does not radiate and rather carries power in the $z$ direction only, but a semi-infinite
TL does radiate, and we show in this section the connection between the finite and semi-infinite case and
how the transition between them occurs. Here, it is important to mention that in some senses a finite matched
TL is very similar to an infinite TL: in both cases there is no reflected wave, but they are very different in
what concerns radiation: the first radiates and the second does not.

In Section~\ref{model} we discuss the radiation resistance and generalize the formula derived in \cite{storer}.

The work is ended with some concluding remarks.

Note: through this work, the phasor amplitudes are RMS values, hence there is no 1/2 in the expressions for power.
Also, it is worthwhile to mention that the results of this work depend on physical sizes relative to the wavelength,
and hence are valid for all frequencies.

\section{Power radiated from a finite TL}
\label{finite}

\subsection{Matched TL}
\label{finitematched}

\begin{figure}[!tbh]
\includegraphics[width=9cm]{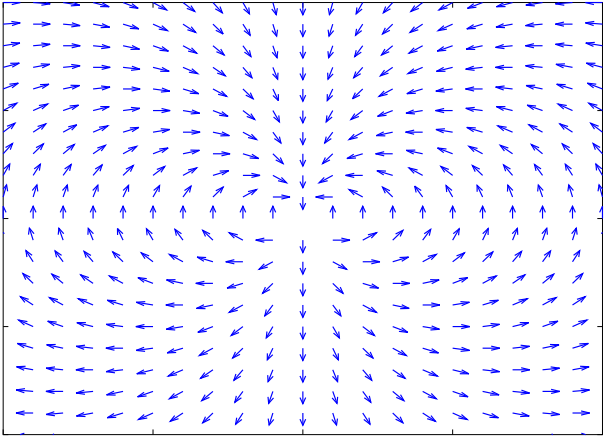}
\caption{The polarization of the $H^+$ field $\boldsymbol{\widehat{\varphi}}\cos\varphi-\boldsymbol{\widehat{\theta}}\sin\varphi$
, according to Eqs.~(\ref{H_explicit}), around $\theta=0$ (``north pole''). The $z$ axis comes toward us from the center
of the plot (at $\theta=0$), $\varphi$ is 0 at the right side and increases counterclockwise.
The polarization is not defined at $\theta=0$, but the fields are 0 at this location.}
\label{polarization}
\end{figure}

We calculate
in this subsection the power radiated form a TL of length $2L$, carrying a forward wave, represented
by the current
\begin{equation}
I(z)=I^+e^{-jkz}
\label{forward_current}
\end{equation}
We set $z_1=-L$ and $z_2=L$ in the expression for
the $z$ directed magnetic vector potential in Eq.~(\ref{A_z_basic_6}) from Appendix~A, and obtain in
spherical coordinates:
\begin{equation}
A_z=\mu_0 G(r)F_{(z)}(\theta,\varphi)
\label{A_z12_1}
\end{equation}
where $G$ is the 3D Green's function defined in (Eq.~(\ref{Green})), and
\begin{equation}
F_{(z)}(\theta,\varphi)=jI^+ 2L kd\sin\theta\cos\varphi\sinc\left[kL(1-\cos\theta)\right]
\label{F_t_p1}
\end{equation}
is the directivity function, and the subscript $(z)$ denotes the contribution from the $z$ directed currents.
The sinc function is defined $\sinc(x)\equiv\sin x/x$. To obtain the far fields (those decaying like $1/r$), the $\boldsymbol{\nabla}$ operator is approximated by
$-jk\mathbf{\widehat{r}}$ and one obtains:
\begin{equation}
\mathbf{H}_{(z)}=\frac{1}{\mu_0}\boldsymbol{\nabla}\times(A_z\mathbf{\widehat{z}})=jkG(r)F_{(z)}(\theta,\varphi)\sin\theta\boldsymbol{\widehat{\varphi}}
\label{H_12}
\end{equation}
and $\mathbf{E}_{(z)}=\eta_0\mathbf{H}_{(z)}\times\mathbf{\widehat{r}}$, where $\eta_0=\sqrt{\mu_0/\epsilon_0}=120\pi\,\Omega$ is the free space impedance.
For the contribution of the $x$ directed end currents we sum the results for the $x$ directed magnetic vector
potential from Eq.~(\ref{A_x12_1}) in Appendix~A (setting $z_1=-L$ and $z_2=L$), obtaining
\begin{equation}
A_{x}=\mu_0 G(r)F_{(x)}(\theta,\varphi)
\label{A_x}
\end{equation}
where
\begin{equation}
F_{(x)}(\theta,\varphi)= 2j I^+d \sin\left[kL(1-\cos\theta)\right]
\label{F_34}
\end{equation}
is the directivity function, and the subscript $(x)$ denotes the contribution from the $x$ directed currents.
The fields from the $x$ directed end currents are calculated, obtaining
\begin{equation}
\mathbf{H}_{(x)}=\frac{1}{\mu_0}\boldsymbol{\nabla}\times(A_x\mathbf{\widehat{x}})=
-jkG(r)F_{(x)}(\cos\theta\cos\varphi\boldsymbol{\widehat{\varphi}} + 
\sin\varphi\boldsymbol{\widehat{\theta}})
\label{H_34}
\end{equation}
and $\mathbf{E}_{(x)}=\eta_0\mathbf{H}_{(x)}\times\mathbf{\widehat{r}}$.
Now summing the fields contributed by the $z$ directed currents with those contributed by the
$x$ directed currents, we obtain $\mathbf{H}=\mathbf{H}_{(z)}+\mathbf{H}_{(x)}$:
\begin{equation}
\mathbf{H}=jkG(r)[\boldsymbol{\widehat{\varphi}}(F_{(z)}\sin\theta-F_{(x)}\cos\theta\cos\varphi)-
                  \boldsymbol{\widehat{\theta}}F_{(x)}\sin\varphi].
\label{H}
\end{equation}
Using Eqs~(\ref{F_t_p1}) and (\ref{F_34}), the explicit expression for the far magnetic field is
\begin{equation}
\mathbf{H}^+=-G(r)2kd I^+ \sin\left[2kL\sin^2(\theta/2)\right][\boldsymbol{\widehat{\varphi}}\cos\varphi-\boldsymbol{\widehat{\theta}}\sin\varphi].
\label{H_explicit}
\end{equation}
We now use the superscript + on all the quantities calculated in this subsection, to denote that they refer to a forward wave.
The far electric field is
\begin{equation}
\mathbf{E}^+=\eta_0\mathbf{H}^+\times\mathbf{\widehat{r}}
\label{E}
\end{equation}
We remark that the polarization of the fields is not well defined at $\theta=0$ (as shown in Figure~\ref{polarization}),
but this is not a problem in this case, because the fields are 0 at this point.

\begin{figure}[!tbh]
\includegraphics[width=9cm]{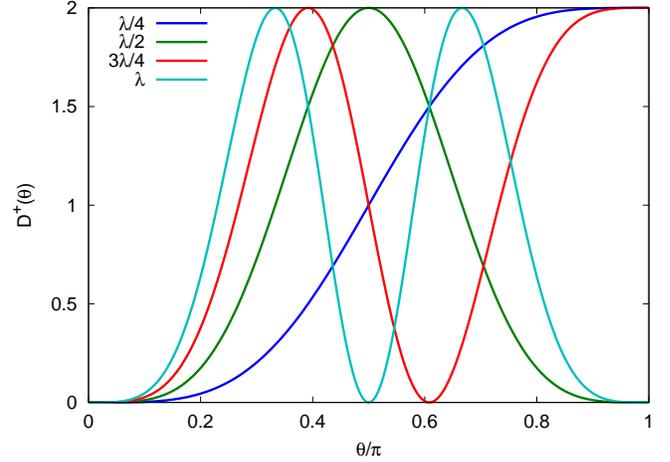}
\caption{$D^+$ as function of $\theta$ for TL $2L=\lambda/4$, $\lambda/2$, $3\lambda/4$ and $\lambda$.}
\label{D_plus}
\end{figure}

The far Poynting vector $\mathbf{S}^+=\mathbf{E}^+\times\mathbf{H}^{+*}$ comes out
\begin{align}
\mathbf{S}^+=\frac{\mathbf{\widehat{r}}\eta_0k^2 d^2|I^+|^2}{4\pi^2r^2}\sin^2\left[2kL\sin^2(\theta/2)\right]
\label{S}
\end{align}
so that the total radiated power is calculated via
\begin{equation}
\int_0^{2\pi}\int_0^{\pi}\sin\theta d\theta d\varphi r^2 \mathbf{\widehat{r}}\cdot\mathbf{S}.
\label{P_rad_1}
\end{equation}
and comes out
\begin{equation}
P_{rad}^+=60\,\Omega |I^+|^2 (kd)^2 \left[1-\sinc(4kL)\right].
\label{P_rad}
\end{equation}
The radiation pattern function is calculated from the radial pointing vector (Eq.~\ref{S}) and the total power
in Eq.~(\ref{P_rad}): $D^+=4\pi r^2 S_r^+/P_{rad}^+$, which comes out
\begin{equation}
D^+(\theta)=2\frac{\sin^2[2kL\sin^2(\theta/2)]}{\left[1-\sinc(4kL)\right]}.
\label{D}
\end{equation}
The function $D^+$ is 0 for $\theta=0$, and its number of lobes increases as the TL length increases.
A one dimensional plot of $D^+$ as function of $\theta$ for different TL lengths is shown in Figure~\ref{D_plus}.

The radiated power relative to the forward wave propagating power ($P^+=|I^+|^2Z_0$) is given by
\begin{equation}
\frac{P_{rad}^+}{P^+}=\frac{60\,\Omega}{Z_0}(kd)^2\left[1-\sinc(4kL)\right],
\label{rel_rad_calc}
\end{equation}
As results from Eq.~(\ref{rel_rad_calc}), the calculation
of the relative radiated power requires the knowledge of two parameters: the separation distance in the
twin lead representation $d$ (relative to the wavelength), and the characteristic impedance of the actual
cross section $Z_0$, and in Appendix~B we show examples of how to calculate this two parameters
for different cross sections.


In the following subsection we generalize the radiation losses for a TL with any termination.

\subsection{Generalization for non matched line}
\label{finitenonmatched}

We generalize here the result (\ref{P_rad}) obtained for the losses of a finite TL carrying
a forward wave to any combination of waves, as follows: 
\begin{equation}
I(z)=I^+e^{-jkz}+I^-e^{jkz}
\label{for_back_current}
\end{equation}
where $I^+$ is the forward wave phasor current, as used in the previous subsection and $I^-$
is the backward wave phasor current, still defined to the right in the ``upper'' line in
Figure~\ref{conf_tl_rad}.

The solution for a backward moving wave (only) on the finite TL, with a current phasor amplitude $I^-$
can be found by first solving for a {\it reversed} $z$ axis in Figure~\ref{conf_tl_rad}, i.e. a $z$ axis going to the
left, replacing in the solution $I^+\rightarrow -I^-$.
But this defines exactly the same configuration for the backward wave, as the original $z$ axis
defined for a forward wave, hence resulting in the same solution in Eqs.~(\ref{H_explicit}) and (\ref{E}).
Now to express the solution for the backward wave in the original coordinates, defined by the
right directed $z$ axis, one has to replace: $\theta\rightarrow\pi-\theta$, $\varphi\rightarrow -\varphi$,
and therefore also $\boldsymbol{\widehat{\theta}}\rightarrow -\boldsymbol{\widehat{\theta}}$ and
$\boldsymbol{\widehat{\varphi}}\rightarrow -\boldsymbol{\widehat{\varphi}}$. Hence, for a backward
wave, the far H field is
\begin{align}
 \mathbf{H}^- = -G(r) 2 kd I^- \sin\left[2kL\cos^2(\theta/2)\right]
 [\boldsymbol{\widehat{\varphi}}\cos\varphi + \boldsymbol{\widehat{\theta}}\sin\varphi],
\label{H_minus}
\end{align}
so that the polarization at $\theta=\pi$ is not defined, and it looks like in
Figure~\ref{polarization}, this time the center of the plot is $\theta=\pi$, and $\varphi$ 
increases clockwise. Again, this is no problem because for this case the fields vanish
at $\theta=\pi$.

For a backward wave only, the radiation pattern is:
\begin{equation}
D^-(\theta)=2\frac{\sin^2[2kL\cos^2(\theta/2)]}{\left[1-\sinc(4kL)\right]},
\label{D_minus}
\end{equation}
which looks like in Figure~\ref{D_plus}, only reflected around $\theta=\pi/2$.

We sum the fields of the forward and backward waves given in Eqs.~(\ref{H_explicit}) and (\ref{H_minus}), obtaining
\begin{align}
&\mathbf{H}= kG(r) 2d \notag \\
&[-\boldsymbol{\widehat{\varphi}}\cos\varphi(I^+\sin\left[2kL\sin^2(\theta/2)\right]+I^-\sin\left[2kL\cos^2(\theta/2)\right])+ \notag \\
& \boldsymbol{\widehat{\theta}}\sin\varphi(I^+\sin\left[2kL\sin^2(\theta/2)\right]-I^-\sin\left[2kL\cos^2(\theta/2)\right])],
\label{HT}
\end{align}


so the far Poynting vector is
\begin{align}
\mathbf{S}=&\mathbf{\widehat{r}}\eta_0\frac{4 (kd)^2}{16\pi^2r^2}\left \{|I^+|^2\sin^2\left[2kL\sin^2(\theta/2)\right]+|I^-|^2\right.\notag \\
& \sin^2\left[2kL\cos^2(\theta/2)\right]+ 2\cos(2\varphi)  \notag \\
& \left. \sin\left[2kL\sin^2(\theta/2)\right]\sin\left[2kL\cos^2(\theta/2)\right]\Re\{I^+I^{-*}\}\right \}
\label{S1_1_f_b}
\end{align}
The interference between the waves does not contribute to the radiated power (because
$\int_0^{2\pi}d\varphi\cos(2\varphi)=0$). The radiated power comes out
\begin{align}
P_{rad}= & 60\,\Omega (kd)^2 \left(|I^+|^2+|I^-|^2\right)\left[1-\sinc(4kL)\right]\equiv \notag \\
        &P^+_{rad}+P^-_{rad},
\label{P_rad1_forw_back}
\end{align}
where $P_{rad}^+$ is given in Eq.~(\ref{P_rad}) and $P^-_{rad}$ is similar, only replace $I^+$ by $I^-$.

\begin{figure}[!tbh]
\includegraphics[width=9cm]{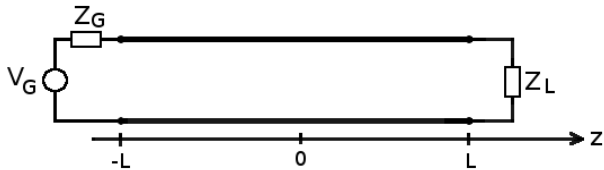}
\caption{TL fed by a generator $V_G$ with an internal impedance $Z_G$, loaded by $Z_L$. The value of $Z_L$ affects
indirectly the radiated power by setting the relation between the forward and backward currents $I^+$ and $I^-$.}
\label{loaded_TL}
\end{figure}

The radiation pattern for the general case of forward and backward waves is $D=4\pi r^2 S_r/P_{rad}$,
where $S_r$ is given in Eq.~(\ref{S1_1_f_b}) and $P_{rad}$ in (\ref{P_rad1_forw_back}). To express $D$
independently of the currents, it is convenient
to consider a general TL circuit in Figure~\ref{loaded_TL},
for which the relation between $I^+$ and $I^-$ is
\begin{equation}
I^-e^{jkL}+\Gamma I^+e^{-jkL}=0,
\label{relation_I_plus_I_minus}
\end{equation}
where $\Gamma\equiv\frac{Z_L-Z_0}{Z_L+Z_0}$. Using (\ref{relation_I_plus_I_minus}), the radiation pattern is
\begin{align}
D(\theta,\varphi)=2\frac{A^2+|\Gamma|^2B^2 - 2AB\cos(2\varphi)\Re\{\Gamma e^{-2jkL}\}}
{\left(1+|\Gamma|^2\right)\left[1-\sinc(4kL)\right]},
\label{D_f_b}
\end{align}
where $A$ and $B$ are abbreviations for:
\begin{equation}
A\equiv\sin\left[2kL\sin^2(\theta/2)\right] \,\,\, B\equiv\sin\left[2kL\cos^2(\theta/2)\right]
\label{A_and_B}
\end{equation}
For a matched TL, $\Gamma=0$, and Eq.~(\ref{D_f_b}) reduces to Eq.~(\ref{D}). We remark that although
the interference between the forward and backward waves does not contribute to the radiated power
(see Eq.~\ref{P_rad1_forw_back}), it distorts the radiation pattern and introduces a $\varphi$
dependence.

As mentioned in the introduction, the ``traveling wave antenna'' presented in
\cite{collin,orfanidis,ramo,jordan,balanis} do not represent transmission lines,
and therefore the radiation patterns in (\ref{D}) or (\ref{D_f_b}) are not comparable with those presented
in the above references. They will be however compared with \cite{Nakamura_2006,Matzner,Guertler}.



In the next section we validate the analytic results obtained in this section, using ANSYS commercial
software simulation and additional published results on radiation losses from TL.

\section{Validation of the analytic results}
\label{validation}
\subsection{Comparison with ANSYS simulation results}
\label{ANSYS}

\begin{figure}[!tbh]
\includegraphics[width=8cm]{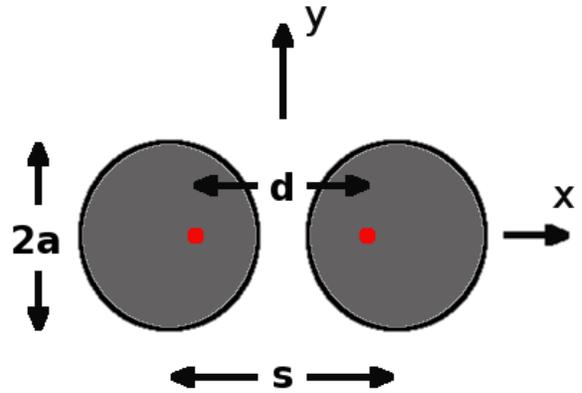}
\caption{Cross section of two parallel cylinders: the distance between the centers is $s=0.02872$,
and the diameters are $2a=0.0203$ wavelengths. The the red points show the current images which 
define the twin lead representation, and the distance between them $d=0.0203$ wavelengths is calculated in
Eq.~(\ref{d}).}
\label{parallel_cylinders}
\end{figure}

We compare in this subsection the relative radiated power from a two-conductor TL (Eq.~\ref{rel_rad_calc})
carrying a forward wave, with the relative radiation losses results obtained from ANSYS commercial software
simulation. For the comparison we use the cross section of two parallel cylinders, for which the analytic solution
is known from image theory \cite{pozar,orfanidis,ramo,jordan}, but also confirmed by Appendix~B. The cross section is shown in
Figure~\ref{parallel_cylinders}. The diameters of the cylinders are $2a=0.0203\lambda$, and the distance between
their centers is $s=0.02872\lambda$, where the wavelength $\lambda=6.25$~cm, corresponding to the frequency of 4.8~GHz.
The distance between the image currents (shown as red points in Figure~\ref{parallel_cylinders}) is the
separation distance $d$ in the twin lead model, given by
\begin{equation}
d=\sqrt{s^2-(2a)^2}=0.0203\lambda,
\label{d}
\end{equation}
so that $(kd)^2=0.016$ is small enough, and the characteristic impedance is
\begin{equation}
Z_0=\frac{\eta_0}{\pi} \ln\left(\frac{d+s}{2a}\right)=105.6\,\Omega
\label{Z0}
\end{equation}
Both analytic results for $d$ and $Z_0$ compare well with those calculated in Appendix~B for
this cross section.

We simulated the configuration in Figure~\ref{parallel_cylinders}
using ANSYS-HFSS commercial software, in the frequency domain, FEM technique.
The box surface enclosing the device constitutes a radiation boundary, implying
absorbing boundary conditions (ABC), used to simulate an open configuration
that allows waves to radiate infinitely far into space. ANSYS HFSS ABC,
absorbs the wave at the radiation boundary, essentially ballooning the
boundary infinitely far away from the structure. The enclosing box surface has
to be located at least a quarter wavelength from the radiating source. For the
frequency of 4.8~GHz we used, the wavelength is 6.25~cm , and we chose the box
sides 7.5~cm in the $x$ and $y$ directions, and the TL length plus 2.5~cm on each
side in the $z$ direction.
For the interface to the device we used lumped ports, which define perfect
H boundaries everywhere on the port plane, so that the E field on the port
plane (outside the conductors) is perpendicular to the conductors.

\begin{figure}[!tbh]
\includegraphics[width=9cm]{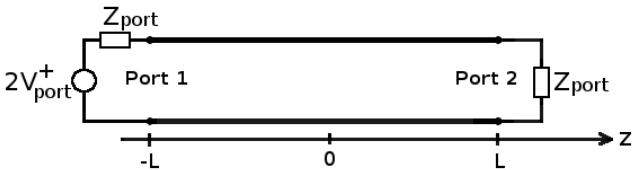}
\caption{Simulation setup for obtaining $2\times 2$ S matrices for different TL lengths.}
\label{two_ports_simulation}
\end{figure}

The simulation setup is shown schematically in Figure~\ref{two_ports_simulation}. The TL is ended at both
sides by lumped ports of characteristic impedance $Z_{port}=50\,\Omega$, but fed only from port 1 by forward
wave voltage $V^+_{port}=1\,V$, so the equivalent Th\'evenin feeding circuit is a generator of $2V^+_{port}$ in series
with a resistance $Z_{port}$.

We obtained from the simulation S matrices defined
for a characteristic impedance of $Z_{port}$ at both ports (which is an arbitrary choice), for different
lengths of the transmission line. By symmetry, the S matrix has the form
\begin{equation}
S=
\left(
\begin{array}{cc}
\Gamma &  \tau  \\
\tau  &    \Gamma 
\end{array}
\right),
\label{S_mat}
\end{equation}
from which one may calculate the ABCD matrix of the TL \cite{pozar, lossless, lossy_MTL, eilat_2014}. We need only the A element from the matrix:
\begin{equation}
A=\frac{1}{2}\left[\tau+(1-\Gamma^2)/\tau\right]
\label{A_D}
\end{equation}
from which we compute the delay angle (or electrical length) of the TL
\begin{equation}
\Theta=\arccos(A)
\label{Theta}
\end{equation}
The real part of $\Theta$ represents the phase accumulated by a forward wave along the TL,
and the imaginary part of $\Theta$ (which is always negative) represents the relative decay of the forward
wave (voltage or current) due to losses (in our case there are only radiation losses) along the TL,
so that $|I^+(L)|=|I^+(-L)|\exp(\text{Im}\{\Theta\})$. Therefore,
the power carried by the forward wave $|P^+(L)|=|P^+(-L)|\exp(2\text{Im}\{\Theta\})$, but for small losses
$|P^+(L)|\simeq |P^+(-L)|(1+2\text{Im}\{\Theta\})$, so that the difference between the input and output values of
$P^+$ (which represent the radiated power $P_{rad}^+$ in Eq.~(\ref{rel_rad_calc})), relative to the (average) power $P^+$ carried by the wave is obtained by
\begin{equation}
\frac{P_{rad}^+}{P^+}=-2 \text{Im}\{\Theta\},
\label{rel_rad_sim}
\end{equation}
where Im is the imaginary part and Im$\{\Theta\}<0$ always.

\begin{figure}[!tbh]
\includegraphics[width=9cm]{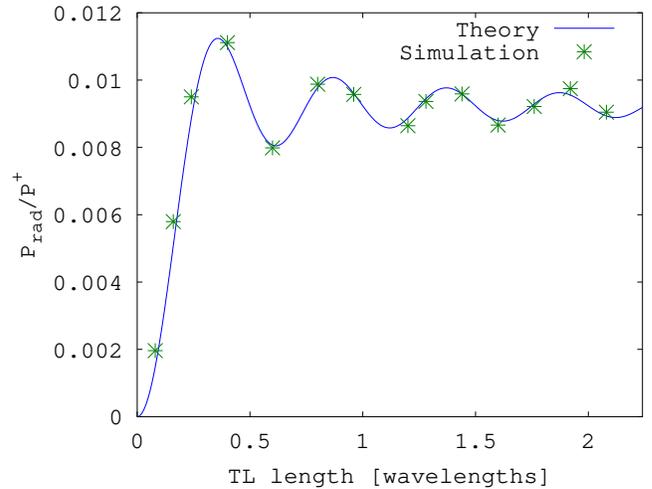}
\caption{Relative radiation losses for a matched TL $P_{rad}/P^+$: comparison between the analytic result in
Eq.~(\ref{rel_rad_calc}) and the simulation result in Eq.~(\ref{rel_rad_sim}) for different TL lengths
in units of wavelengths.}
\label{rad_losses_comparison}
\end{figure}

In Figure~\ref{rad_losses_comparison} and Table~\ref{rad_losses_comparison_t} we compare the analytic result
in Eq.~(\ref{rel_rad_calc}) with the result obtained from simulation Eq.~(\ref{rel_rad_sim}), for a fixed frequency of 4.8~GHz and different
TL lengths.
\begin{table}[!tbh]
\centering
\caption{The numerical data from Figure~\ref{rad_losses_comparison} and the relative error.}
\label{rad_losses_comparison_t}
\begin{tabular}{|c|c|c|c|}
\hline
TL length          &    simulation (Eq.~\ref{rel_rad_sim})   &  theoretical (Eq.~\ref{rel_rad_calc}) & \% error \\ \hline
        0.08  &   0.001963  &   0.001479  &    32.76  \\ \hline
        0.16  &   0.005796  &   0.005079  &    14.12  \\ \hline
        0.24  &   0.009502  &   0.008851  &     7.35  \\ \hline
         0.4  &   0.011115  &   0.010982  &     1.21  \\ \hline
         0.6  &   0.007987  &   0.008069  &    -1.02  \\ \hline
         0.8  &   0.009878  &   0.009774  &     1.06  \\ \hline
        0.96  &   0.009575  &   0.009603  &    -0.28  \\ \hline
        1.12  &   0.008198  &   0.008579  &    -4.44  \\ \hline
         1.2  &   0.008646  &   0.008874  &    -2.57  \\ \hline
        1.28  &   0.009366  &   0.009445  &    -0.84  \\ \hline
        1.44  &   0.009589  &   0.009583  &     0.05  \\ \hline
         1.6  &   0.008665  &   0.008797  &    -1.50  \\ \hline
        1.76  &   0.009216  &   0.009286  &    -0.75  \\ \hline
        1.92  &   0.009746  &   0.009557  &     1.98  \\ \hline
        2.08  &   0.009046  &   0.008936  &     1.23  \\ \hline
\end{tabular}
\end{table}
We see that the simulation confirms well the theoretical result, with an average absolute relative
error of 4.75\%. The biggest relative errors are at the short TL, where the relative radiation losses
are low, and hence more difficult to reproduce accurately with the simulation. For example, if we exclude
the shortest TL length of 0.08$\lambda$ from the comparison, the average absolute relative error is 2.75\%, or if
we exclude the 2 shortest points of 0.08 and 0.16$\lambda$ from the comparison, it drops to 1.87\%.

\subsection{Comparison with \cite{Manneback}}
\label{comp_manneback}

In 1923 Manneback \cite{Manneback} published a paper ``Radiation from transmission lines'' which
calculated the power radiated by two thin wires of length $l$ (equivalent to our $2L$), and separation
$d$, in resonance, having open terminations. The author considered the current
\begin{equation}
I_m \cos(k_nz)\sin(\omega_nt)
\label{I_manneback}
\end{equation}
where $k_n=\pi n/l$ and $\omega_n=ck_n$ for odd $n$ (as defined in Eq.~3 of \cite{Manneback}).
Note that the time dependence has been explicitly written in \cite{Manneback} and the calculations
have been done in time domain, but using a fixed frequency, hence they are completely equivalent to our
phasor calculations. The result for the radiated power is given in Eq.~11 of \cite{Manneback}, rewritten
here for convenience
\begin{equation}
P_{rad}=15\,\Omega (kd)^2 I_m^2
\label{P_rad_manneback}
\end{equation}
We compare this with our result (\ref{P_rad1_forw_back}). $kl=n\pi$ is in our notation $kL=n\pi/2$
hence the sinc function in Eq.~(\ref{P_rad1_forw_back}) results 0.
This resonant case implies $I^+=I^-$, so the general current in Eq.~(\ref{for_back_current}) reduces to
\begin{equation}
2I^+\cos(kz)
\label{I_resonant}
\end{equation}
and Eq.~(\ref{P_rad1_forw_back}) results in $120\,\Omega(kd)^2|I^+|^2$.

The value $I_m$ in
Eq.~(\ref{I_manneback}) is the amplitude of the current, i.e. the RMS value times $\sqrt{2}$.
We use RMS values (as mentioned in the introduction) hence the equivalence between  
Eq.~(\ref{I_manneback}) and Eq.~(\ref{I_resonant}) is by setting $I_m=2\sqrt{2}|I^+|$,
and using this equivalence, Eq.~(\ref{P_rad1_forw_back}) reduces {\it exactly} to
Eq.~(\ref{P_rad_manneback}).

To be mentioned that our results are general, covering all cases of terminations, or
any combination of waves, and we showed in this subsection how our general result reduces
correctly to the result for a private case of resonance.

\subsection{Comparison with \cite{storer}}
\label{storer_comparison}

In 1951 J. E. Storer and R. King, published the paper ``Radiation Resistance of a Two-Wire Line'', in which
they calculated the radiation resistance of a twin lead TL loaded by an arbitrary load, i.e. carrying an arbitrary
combination of forward and backward waves, shown schematically in Figure~\ref{loaded_TL}.
Defining the complex reflection coefficient $\Gamma\equiv\frac{Z_L-Z_0}{Z_L+Z_0}$, the relation between $I^+$
and $I^-$ is given in Eq.~(\ref{relation_I_plus_I_minus}).
The current at the generator side is:
\begin{equation}
I(-L)=I^+e^{jkL}+I^-e^{-jkL},
\label{I_gen}
\end{equation}
and the radiation resistance is defined by
\begin{equation}
r_{\text{rad}}\equiv P_{\text{rad}}/|I(-L)|^2,
\label{r_rad_def}
\end{equation}
where $P_{\text{rad}}$ is given in Eq.~(\ref{P_rad1_forw_back}). Using Eqs.~(\ref{relation_I_plus_I_minus}), (\ref{I_gen}) and
(\ref{r_rad_def}) we obtain
\begin{equation}
r_{\text{rad}}=60\,\Omega (kd)^2 \left[1-\sinc(4kL)\right]\frac{1+|\Gamma|^2}{|1-\Gamma e^{-4jkL}|^2},
\label{r_rad_any_waves}
\end{equation}
which is identical to Eq.~(5) in \cite{storer}, after setting the ohmic attenuation $\alpha$ to 0, and use
the identity $\arctan(x)=\frac{1}{2}\ln\frac{1+x}{1-x}$ and the definition of the $\cosh$ function.

One remarks that $r_{\text{rad}}$ in Eq.~(\ref{r_rad_any_waves}) goes to infinity if $|\Gamma|=1$ and
$4kL-\angle\Gamma=2\pi n$ ($n$ integer). This lacuna will be fixed in Section~\ref{model}.

In the private case of a matched TL, using the radiated power in Eq.~(\ref{P_rad}) divided by $|I^+|^2$, or alternatively
setting $\Gamma=0$ in Eq.~(\ref{r_rad_any_waves}) results in:
\begin{equation}
r_{\text{rad}}=60\,\Omega (kd)^2 \left[1-\sinc(4kL)\right],
\label{r_rad}
\end{equation}
which is identical to the case shown in Eq.~(6) in \cite{storer}, after setting the attenuation $\alpha$ to 0.

\subsection{Comparison with \cite{Bingeman_2001}}

Another comparison is with Bingeman's work from 2001 \cite{Bingeman_2001}, in which the
method of moments (MoM) has been used to calculate the radiation from two thin wires of diameter
$2a=5$~mm, length $2L=10$~m and separated at a distance of $d=1$~m. The characteristic impedance
is given by Eq.~(\ref{Z0}) (but given $d\gg a$ one may use $s=d$, see Eq.~(\ref{d})) and results
in $Z_0=720\,\Omega$, as calculated at the beginning of \cite{Bingeman_2001}. In absence of
other losses, the author derived the radiated power as the difference between the power carried
by the TL and the power reaching the load.

\begin{figure}[!tbh]
\includegraphics[width=9cm]{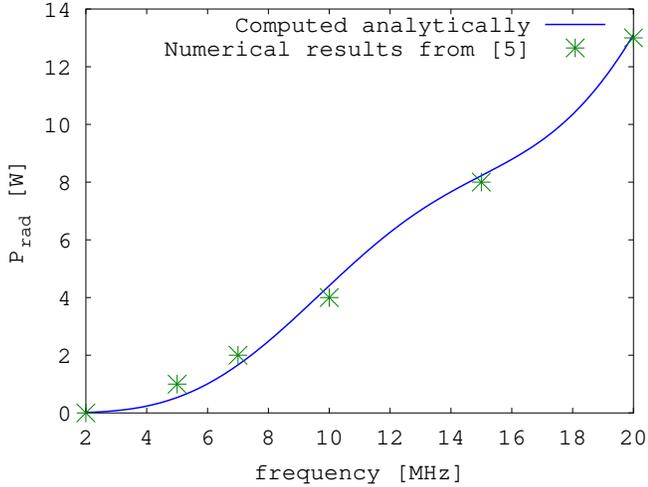}
\caption{Comparison between the radiated power for a matched line from Table~1 of \cite{Bingeman_2001}
with our results for a matched line in Eq.~(\ref{P_rad}) for different frequencies.}
\label{bingerman_matched}
\end{figure}

The first calculation is the power radiated by a matched TL, fed by a power of 1000~W, for frequencies
$f=2$, 5, 7, 10, 15 and 20~MHz. For the power of 1000~W, the RMS value of the forward current to set in
Eq.~(\ref{P_rad}) is $|I^+|=\sqrt{1000/Z_0}=$1.1785~A, and we use $k=2\pi f/c$ for the above frequencies.
The numerical results for this case are given in Table~1 of \cite{Bingeman_2001}, and we compare those
results to ours, in Figure~\ref{bingerman_matched} and Table~\ref{comparison_matched_line}.
\begin{table}[!tbh]
\centering
\caption{The numerical data from Figure~\ref{bingerman_matched} and the relative error.}
\label{comparison_matched_line}
\begin{tabular}{|c|c|c|c|}
\hline
Frequency [MHz]          &    \cite{Bingeman_2001}  &  theoretical (Eq.~\ref{P_rad}) & \% error \\ \hline
2   & 0  & 0.0165  & -100        \\ \hline
5   & 1  & 0.5359  &   87        \\ \hline
7   & 2  & 1.664   &   20        \\ \hline
10  & 4  & 4.411   &  -9.32      \\ \hline
15  & 8  & 8.225   &  -2.73      \\ \hline
20  & 13 & 13.11   &  -0.84      \\ \hline
\end{tabular}
\end{table}
We see a good match for the high frequencies (big electric delay), and it deteriorates at small
electric delays. But the result 0 for the frequency of 2~MHz is clearly incorrect, so we may understand
that the accuracy of the results in \cite{Bingeman_2001} is low at small electric delays, for
which the relative radiated power is small.

\begin{figure}[!tbh]
\includegraphics[width=9cm]{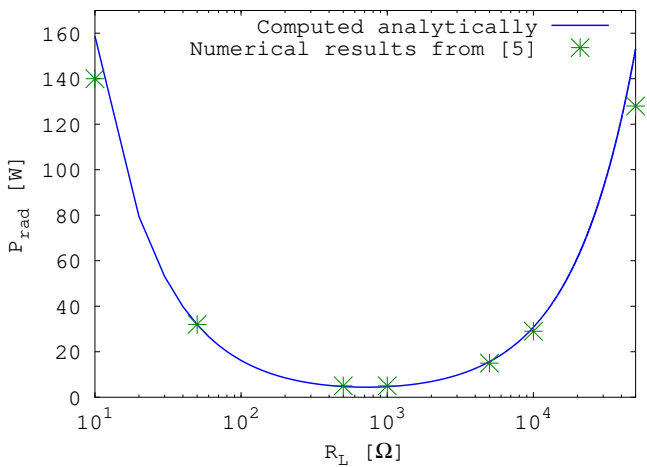}
\caption{Comparison between the radiated power for a non matched line from Table~2 of \cite{Bingeman_2001}
with our results for a non matched line in Eq.~(\ref{P_rad1_forw_back}) for different load resistances.}
\label{bingerman_non_matched}
\end{figure}

Another calculation in \cite{Bingeman_2001} is for a non matched TL, with end loads
$R_L$=10, 50, 500$\,\Omega$, 1, 5, 10, and 50$~k\Omega$, all cases at frequency 10~MHz, carrying
a net power of 1000~W. We compare those results with the results of Eq.~(\ref{P_rad1_forw_back}).
First $k=2\pi f/c=0.2094$~[1/m] is fixed, and we calculate for each load resistance $R_L$
\begin{equation}
|\Gamma|=\left|\frac{R_L-Z_0}{R_L+Z_0}\right|,
\label{absgamma}
\end{equation}
from which the forward power values for each case are given by
\begin{equation}
P^+=\frac{P}{1-|\Gamma|^2}=\frac{1000}{1-|\Gamma|^2}.
\label{P_plus}
\end{equation}
The forward current values for each case are given by $|I^+|=\sqrt{P^+/Z_0}$
and the backward current values for each case are given by $|I^-|=|\Gamma||I^+|$. Setting the
values in Eq.~(\ref{P_rad1_forw_back}), we compare the results of \cite{Bingeman_2001} for
the unmatched line at 10MHz (Table~2 in \cite{Bingeman_2001}), with the results of 
Eq.~(\ref{P_rad1_forw_back}) in Figure~\ref{bingerman_non_matched} and Table~\ref{comparison_non_matched_line}.
\begin{table}[!tbh]
\centering
\caption{The numerical data from Figure~\ref{bingerman_non_matched} and the relative error.}
\label{comparison_non_matched_line}
\begin{tabular}{|c|c|c|c|}
\hline
Load $R_L$ [$\Omega$]    &    \cite{Bingeman_2001}  &  theoretical (Eq.~\ref{P_rad1_forw_back}) & \% error \\ \hline
10   &  140 &  158.83 &  -11.85   \\ \hline
50   &   32 &   31.91 &    0.28   \\ \hline
500  &    5 &    4.70 &    6.38   \\ \hline
1k   &    5 &    4.65 &    7.53   \\ \hline
5k   &   15 &   15.63 &   -4.03   \\ \hline
10k  &   29 &   30.79 &   -5.81   \\ \hline
50k  &  128 &  153.19 &  -16.44   \\ \hline
\end{tabular}
\end{table}
The match between the results is good, except for the first and last cases, in which
the radiated power is big and approaches the order of magnitude of the net power (1000~W).
This may be due to the limitations of the current theory to small losses that almost
do not affect the basic electromagnetic solution (see Introduction).

\subsection{Comparison with \cite{Nakamura_2006}}

In 2006 Nakamura et. al. published the paper ``Radiation Characteristics of a Transmission Line
with a Side Plate'' \cite{Nakamura_2006} which intends to reduce radiation losses from a twin lead TL using
a side plate. The side plate is a perfect conductor put aside the transmission line, to create
opposite image currents, and hence reduce the radiation.

The authors first derived the radiation from a TL without the side plate, obtaining
an integral (Eq.~20 in \cite{Nakamura_2006}) which they computed numerically. The numerical
integration result is shown in Figure~6 of \cite{Nakamura_2006}, where the solid line represents
the free space case.

\begin{figure}[!tbh]
\includegraphics[width=9cm]{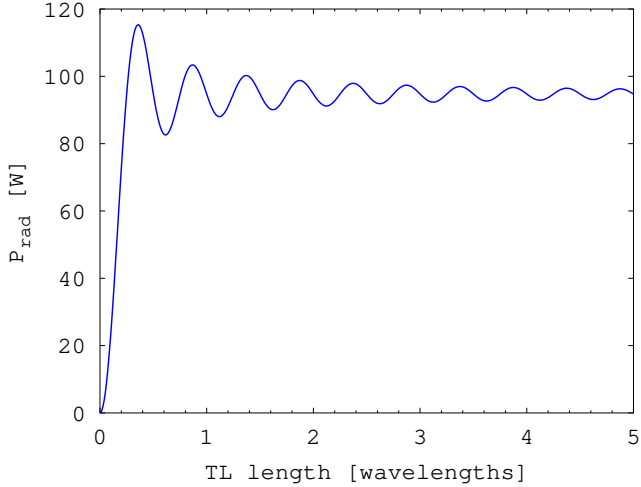}
\caption{Recalculation of the solid line in Figure~6 of \cite{Nakamura_2006}, using Eq.~(\ref{P_rad}).}
\label{comparison_nakamura}
\end{figure}

We compare our analytic result in Eq.~(\ref{P_rad}), with the numerical result shown in
Figure~6 of \cite{Nakamura_2006}. First, $I_0$ in ~\cite{Nakamura_2006} is a forward
current, and from Eq.~(19) in ~\cite{Nakamura_2006}, it is evident that they used RMS values.
They used $I_0=1$A, hence we set $|I^+|=1$A in Eq.~(\ref{P_rad}). $2h$ is the distance
between the conductors in ~\cite{Nakamura_2006}, equivalent to $d$ in this work, and they
used $h\lambda=0.1$, therefore $(kd)^2=(4\pi h/\lambda)^2=1.5791$ in Eq.~(\ref{P_rad}),
so that the total radiated power for the case displayed in Figure~6 of \cite{Nakamura_2006} is
\begin{align}
P_{rad}=&60\times 1\times 1.5791 \left[1-\sinc(4kL)\right]= \notag \\
       &94.746 [W] \left[1-\sinc\left(\frac{2L}{\lambda}4\pi\right)\right],
\label{P_rad_nakamura}
\end{align}
and we wrote the argument of the sinc function in terms of $2L/\lambda$, i.e. the TL length in
wavelengths. This result is displayed in Figure~\ref{comparison_nakamura}, in which we show
the radiated power as function of the TL length in wavelengths. The authors did not supply the
numerical data to reproduce Figure~6 of \cite{Nakamura_2006}, and we did not want to copy 
the figure into this work for comparison, but we checked very carefully
that indeed our calculation shown in Figure~\ref{comparison_nakamura} {\it completely overlaps} the solid line in
Figure~6 of \cite{Nakamura_2006}.

\begin{figure}[!tbh]
\includegraphics[width=4.3cm]{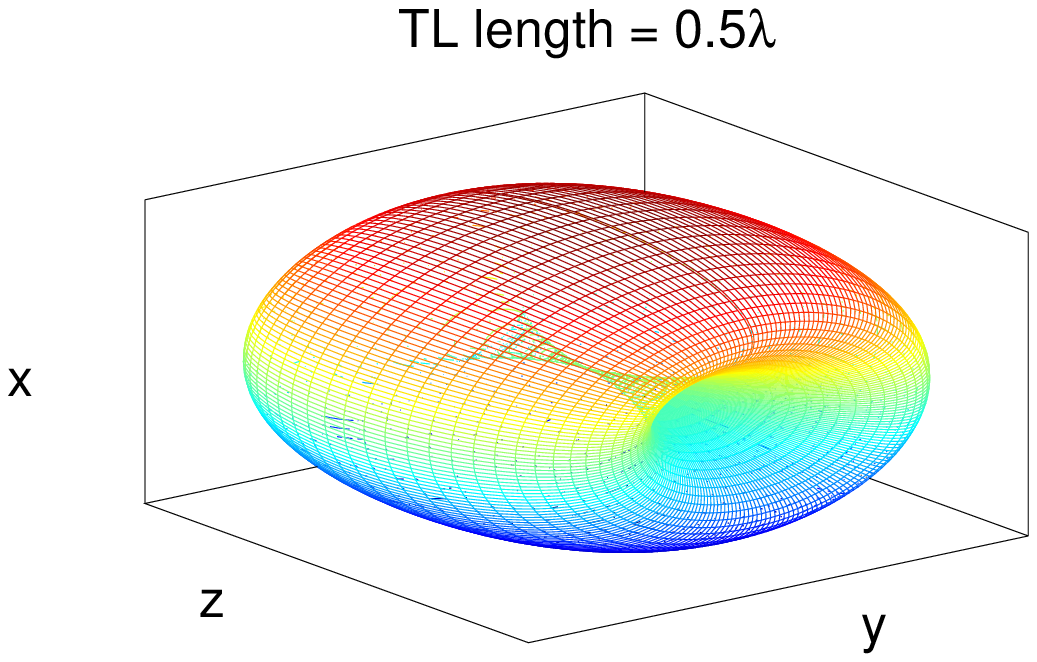}
\includegraphics[width=4.3cm]{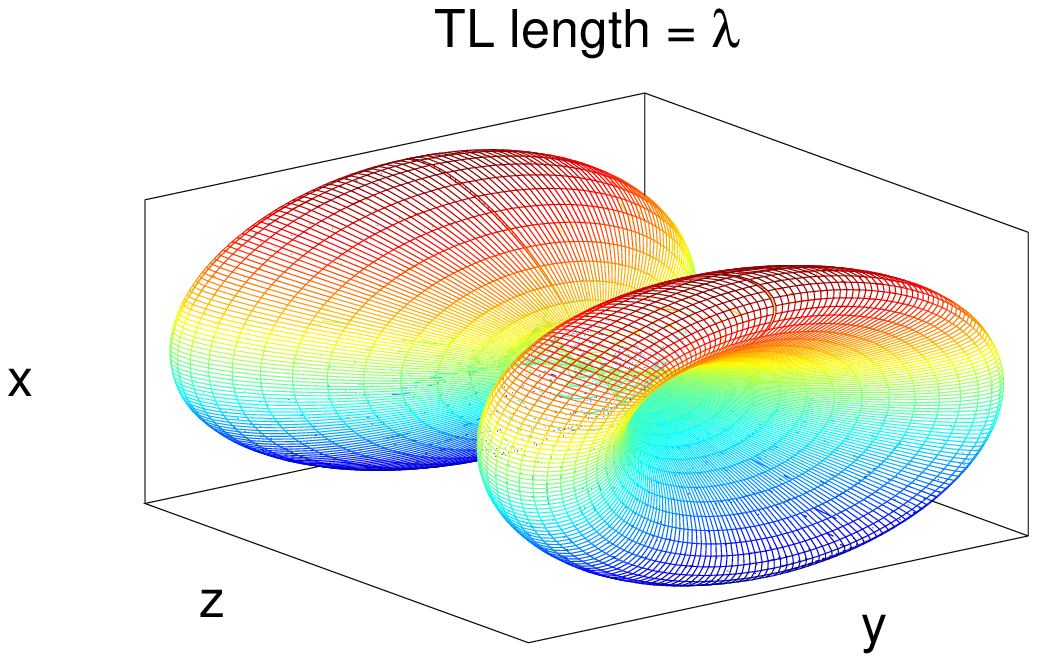}
\includegraphics[width=4.3cm]{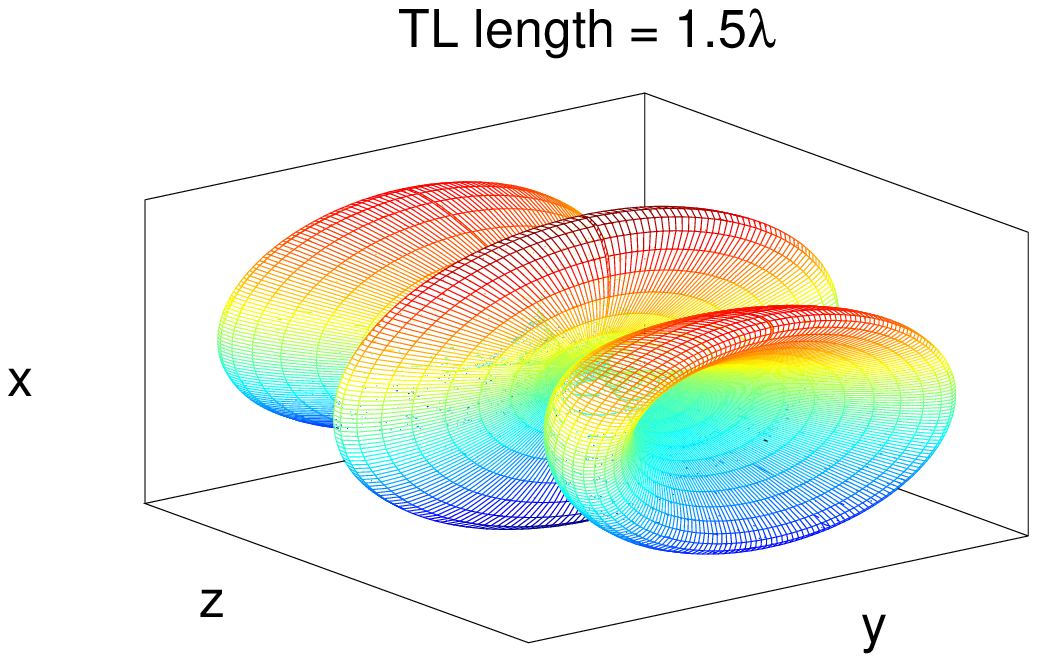}
\includegraphics[width=4.3cm]{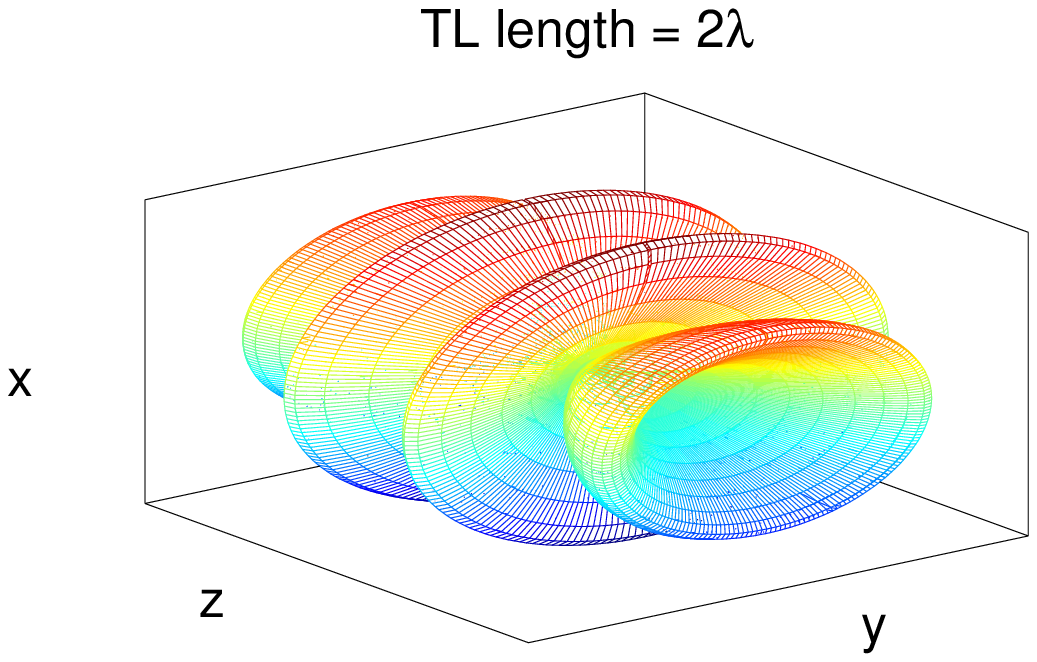}
\caption{Radiation pattern $D^+$ calculated from Eq.~(\ref{D}) for the cases of TL lengths $n\lambda/2$, for n=1 to 4.
They are identical to the parallel cases shown in Figure~5, panel(a) of \cite{Nakamura_2006}. Note
that the definitions of the $x$ and $z$ axes are swapped in \cite{Nakamura_2006} compared to our definitions, we
therefore showed them in an orientation which makes the comparison easy (i.e. our $z$ axis is oriented in the
plots in the same direction as their $x$ axis).
We remark that the pattern for the case $0.5\lambda$ is quite similar to this of a dipole antenna, because a short TL behaves
similar to a small magnetic loop, i.e. a magnetic dipole.}
\label{rad_patt}
\end{figure}

We compare as well the radiation patterns obtained in \cite{Nakamura_2006} with ours (Eq.~\ref{D})
in Figure~\ref{rad_patt}.
We remark that $D^+(\theta)$ is not symmetric around $\theta=\pi/2$ in general (see Figure~\ref{D_plus}), but
for the cases $kL=n\pi/2$ (integer $n$), i.e. the TL length is a multiple integer of half wavelength, displayed
in Figure~\ref{rad_patt}, $D^+(\theta)$ is symmetric around $\theta=\pi/2$, because $\sin^2(n\pi/2+x)=\sin^2(n\pi/2-x)$
for any $x$. The radiation patterns in Figure~\ref{rad_patt} are {\it identical} to the parallel cases shown in Figure~5,
panel(a) of \cite{Nakamura_2006}.

It is worthwhile to remark that the radiation pattern (Eq.~\ref{D}) does not depend on the distance between
the conductors $d$ (or $2h$ in \cite{Nakamura_2006}), hence the annotation of $h/\lambda=0.1$ in Figure~5 of 
\cite{Nakamura_2006} is redundant, and probably has been added to the caption because the authors computed
the radiation patterns numerically for $h/\lambda=0.1$, without deriving an analytic expression.

\subsection{Comparison with \cite{Matzner,Guertler}}
\label{matzner_comparison}

\begin{figure}[!tbh]
\includegraphics[width=9cm]{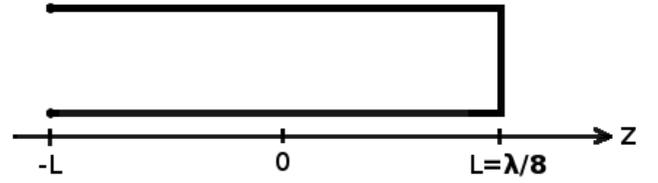}
\caption{Transmission line of length $\lambda/4$, open at one termination, and shorted at the other, represents
a ``U'' shaped antenna.}
\label{u_shaped}
\end{figure}

References \cite{Matzner,Guertler} analyze the radiation from a ``U'' shaped antenna (see Figure~\ref{u_shaped})
and showed that its radiation pattern is uniform.
Using $\Gamma=-1$ (shorted termination) and $kL=\pi/4$ the radiation pattern in Eq.~(\ref{D_f_b}) is:
\begin{align}
D=\sin^2\left[(\pi/2)\sin^2(\theta/2)\right]+\sin^2\left[(\pi/2)\cos^2(\theta/2)\right]=1.
\label{D_U_shaped}
\end{align}
Using $(\pi/2)\cos^2(\theta/2)=(\pi/2)-(\pi/2)\sin^2(\theta/2)$ one easily remarks that (\ref{D_U_shaped})
is 1, describing uniform radiation, as mentioned in \cite{Matzner,Guertler}. This does not contradict the
``hairy-ball'' theorem \cite{Brouwer}, because this theorem states that any {\it real} tangential field must be
0 at least at one point on a sphere, and by {\it real} one means: having the {\it same phase} everywhere.

And indeed the separate fields associated with $I^+$ and $I^-$ (each one ``real'' in the above sense)
given in Eqs.~(\ref{H_explicit}) and (\ref{H_minus})
are 0 at points $\theta=0$ and $\pi$ respectively. From Eq.~(\ref{relation_I_plus_I_minus}), the relation between
the forward and backward wave is $I^-=-jI^+$, and after summing the fields by setting this relation into Eq.~(\ref{HT}),
the far magnetic field is
\begin{align}
&\mathbf{H}= G(r) 2 kd I^+\notag \\
&[-\boldsymbol{\widehat{\varphi}}\cos\varphi e^{-j(\pi/2)\cos^2(\theta/2)} + \boldsymbol{\widehat{\theta}}\sin\varphi e^{j(\pi/2)\cos^2(\theta/2)} ],
\label{H_U_shaped}
\end{align}
which has a constant amplitude everywhere, but a changing phase, which cannot be factored out to get a ``real'' field.
This complex field manifests a linear polarization at $\theta=0$ and $\pi$, circular polarization at
$\theta=\pi/2$ and $\varphi$ multiple of $\pi/4$, and elliptic elsewhere, compare with \cite{Matzner,Guertler}. 

\section{Infinite and semi-infinite TL analysis}
\label{inf_and_semiinf}

As we know, there are no infinite or semi-infinite TL in reality, but the literature considers
those kind of TL as limiting cases, and as we shall see, the analysis of the infinite and
semi-infinite TL supplies an additional insight and validation of the results obtained in the
Section~II, as shown in the following subsections.

Certainly those cases can be considered only in absence of other losses, like ohmic or dielectric,
for which infinite TL have infinite losses. As one remarks, the radiation losses of finite TL reach
an asymptotic value for long TL, so that one may expect that infinite or semi-infinite TL do not
radiate an infinite power.

\subsection{Infinite TL}


For an infinite TL, carrying a forward wave, we set $z_1=-L$ and $z_2=L$ in the result (\ref{A_z_basic_6}) and
considering $L\rightarrow\infty$ (i.e. for finite $z$ and $\rho$, $L\gg|z|,\rho$) we obtain
\begin{equation}
A_z=\frac{\mu_0 I^+}{4\pi} d \cos\varphi\frac{2 e^{-jkz}}{\rho}.
\label{A_z_inf3}
\end{equation}
Certainly, we do not have in this case $x$ directed currents, so we obtain from Eq.~(\ref{A_z_inf3}):
\begin{equation}
\mathbf{H}=\frac{1}{\mu_0}\boldsymbol{\nabla}\times\mathbf{A}=\frac{e^{-jkz}}{2\pi} \frac{I^+ d}{\rho^2} [-\widehat{\boldsymbol{\rho}}\sin\varphi + \widehat{\boldsymbol{\varphi}}\cos\varphi]
\label{H_inf}
\end{equation}
and
\begin{equation}
\mathbf{E}=\frac{1}{j\omega\epsilon_0}\boldsymbol{\nabla}\times\mathbf{H}=\eta_0\frac{e^{-jkz}}{2\pi} \frac{I^+ d}{\rho^2} [\widehat{\boldsymbol{\rho}}\cos\varphi + \widehat{\boldsymbol{\varphi}}\sin\varphi],
\label{E_inf}
\end{equation}
which are the static $\mathbf{H}$ and $\mathbf{E}$ fields multiplied by the forward wave propagation factor $e^{-jkz}$.

We remark that in Appendix~A we considered the far field ($k\rho\gg 1$), so the fields in Eqs.~(\ref{H_inf})
and (\ref{E_inf}) are correct far from the TL, and their diverging at $\rho=0$ is an artifact of this far field
approximation. But even in the far field, writing them in spherical coordinates so that $\rho=r\sin\theta$,
the fields decay like $1/r^2$ and there is no ``radiating'' term decaying like $1/r$.

This is also evident from the Poynting vector:
\begin{equation}
\mathbf{S}=\mathbf{E}\times\mathbf{H}^*=\eta_0\frac{1}{4\pi^2} \frac{|I^+|^2 d^2}{\rho^4} \widehat{\mathbf{z}},
\label{S_inf}
\end{equation}
which is only in the $z$ direction, representing the power carried by the TL.

Given the fact that the fields in Eqs.~(\ref{H_inf}) and (\ref{E_inf}) decay much faster than
radiating fields, hence are negligible relative to them far from the TL, it is convenient to
define a typical distance $\rho_0$ from the TL, so that
\begin{align}
&\text{Static near field is dominant if}\,\,\rho<\rho_0 \notag \\
&\text{Radiation field is dominant if}\,\,\rho>\rho_0 .
\label{rho_0}
\end{align}
There are no radiation fields in this subsection, but the relation (\ref{rho_0}) will be referred
to in the next subsection, analyzing semi-infinite TL.

Another way of understanding $\rho_0$ is
by integrating the Poynting vector to obtain the forward power $P^+$
\begin{equation}
P^+=\iint_{-\infty}^{\infty}dx dy\,\mathbf{S}\cdot\widehat{\mathbf{z}},
\label{P_plus_inf}
\end{equation}
and here one has to use the exact fields in the expression for $\mathbf{S}$ (not the far fields in
Eqs.~(\ref{H_inf}) and (\ref{E_inf})). To obtain $P^+$ with a ``reasonable'' required accuracy, one does
not need to integrate to infinity, but rather
\begin{equation}
P^+\simeq \int_{0}^{2\pi}d\varphi \int_{0}^{\rho_0} d\rho\, \rho \,\mathbf{S}\cdot\widehat{\mathbf{z}},
\label{P_plus_inf_appr}
\end{equation}
so that $\rho_0$ is the radial distance from the TL (in cylindrical coordinates) within which the
near fields are significant.


\subsection{Semi-infinite TL}

\begin{figure}[!tbh]
\includegraphics[width=9cm]{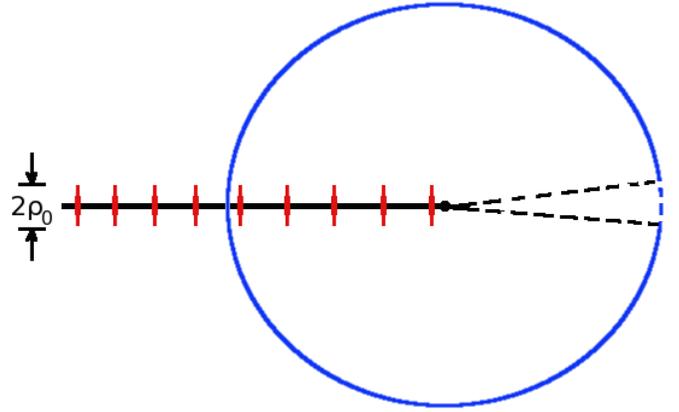}
\caption{Semi-infinite TL from $z=-\infty$ to the center of coordinates at $z=0$. The blue circle represents
the wave front of the outgoing spherical wave radiation in the second part of Eq.~(\ref{A_z_semiinf_minus_infty_to_0})
and the red wave fronts represent the near plane wave field in the first part of
Eq.~(\ref{A_z_semiinf_minus_infty_to_0}). The near plane wave and spherical wave cancel each other in
the paraxial region $z>0$ and $\rho<\rho_0$, see Eq.~(\ref{spherical_on_axis}). The spherical wave is
shown dashed blue in the canceling region, which occurs within a cone $\Delta\theta=\rho_0/r$
(dashed black line). The cone gets narrower as the distance from the center of coordinates $r$ increases.}
\label{semi_infinite_RT}
\end{figure}

\begin{figure}[!tbh]
\includegraphics[width=9cm]{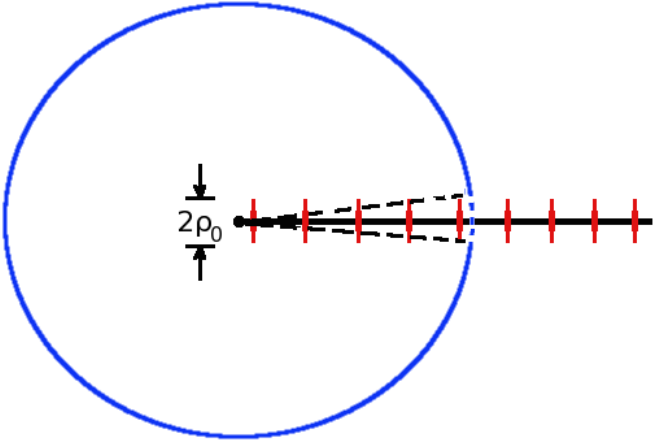}
\caption{Semi-infinite TL from the center of coordinates at $z=0$ to $z=\infty$. The blue circle represents
the wave front of the outgoing spherical wave radiation (Eq.~(\ref{A_z_semiinf_0_to_infty})), but behaves
in the paraxial region $z>0$ and $\rho<\rho_0$, like the near plane wave in the first part of
Eq.~(\ref{A_z_semiinf_minus_infty_to_0}), according to Eq.~(\ref{spherical_on_axis}). Therefore, in this paraxial
region the spherical wave front (dashed blue) does not represent radiation, but rather the near plane wave.
Like in Figure~\ref{semi_infinite_RT}, this paraxial region is within the cone $\Delta\theta=\rho_0/r$
(dashed black line), which gets narrower as the distance from the center of coordinates $r$ increases.}
\label{semi_infinite_LT}
\end{figure}

We analyze here a semi-infinite TL carrying a forward wave. The TL can be either from $z=-\infty$ to 0
(Figure~\ref{semi_infinite_RT}) or from $z=0$ to $\infty$ (Figure~\ref{semi_infinite_LT}). We note that
in both cases we have to consider also the contribution of the $x$ directed current at the termination at $z=0$.

For the first case we set $z_1=-L$, $z_2=0$ in Eq.~(\ref{A_z_basic_6}) and taking $L\rightarrow\infty$ we obtain
\begin{equation}
A_{z\,\, RT}=\frac{\mu_0 I^+}{4\pi} d \cos\varphi\left[\frac{2 e^{-jkz}}{\rho}-\frac{\rho}{r-z}\frac{e^{-jkr}}{r}\right]
\label{A_z_semiinf_minus_infty_to_0}
\end{equation}
and from Eq.~(\ref{A_x12_1}) for $z_2=0$:
\begin{equation}
A_{x\,\, RT}=-\mu_0 I^+ d \frac{e^{-jkr}}{4\pi r},
\label{A_x_semiinf_minus_infty_to_0}
\end{equation}
while for the second case we set $z_1=0$, $z_2=L$ in Eq.~(\ref{A_z_basic_6}) and taking $L\rightarrow\infty$ we obtain
\begin{equation}
A_{z\,\, LT}=\frac{\mu_0 I^+}{4\pi} d \cos\varphi \left[\frac{\rho}{r-z}\frac{e^{-jkr}}{r}\right]
\label{A_z_semiinf_0_to_infty}
\end{equation}
and from Eq.~(\ref{A_x12_1}) for $z_1=0$:
\begin{equation}
A_{x\,\, LT}=\mu_0 I^+ d \frac{e^{-jkr}}{4\pi r},
\label{A_x_semiinf_0_to_infty}
\end{equation}
where the subscripts RT and LT mean ``right terminated'' and ``left terminated'' TL, respectively.
Looking at the RT configuration in Eq.~(\ref{A_z_semiinf_minus_infty_to_0}), we see that it includes also the near field
expression of the infinite TL from Eq.~(\ref{A_z_inf3}) for all $z$ in spite of the fact that the TL
is in the region $z<0$ and terminates at $z=0$. This is explained by the fact that for $z>0$ and small
$\rho$ (typically $\rho<\rho_0$, see (\ref{rho_0})), $r\approx z+\frac{\rho^2}{2z}$, so that
$r-z\approx\frac{\rho^2}{2z}$, resulting in
\begin{equation}
\left.\frac{\rho}{r-z}\frac{e^{-jkr}}{r}\right|_{\substack{z>0 \\ \rho<\rho_0}} \simeq \frac{2 e^{-jkz}}{\rho},
\label{spherical_on_axis}
\end{equation}
which means that the spherical wave in the second part of Eq.~(\ref{A_z_semiinf_minus_infty_to_0}) describes
radiation everywhere except in a cone around $\theta=0$ where it is canceled by the near plane wave in
the first part of Eq.~(\ref{A_z_semiinf_minus_infty_to_0}), see Figure~\ref{semi_infinite_RT}.

For the LT configuration the spherical wave in Eq.~(\ref{A_z_semiinf_0_to_infty}) represents radiation
except inside a cone around $\theta=0$, where it equals the near plane wave in the first part of
Eq.~(\ref{A_z_semiinf_minus_infty_to_0}) (according to Eq.~(\ref{spherical_on_axis})), which does not have
radiating fields, see Figure~\ref{semi_infinite_LT}.

So to calculate the radiated power $P_{rad}^+$, one may either use the second part of Eq.~(\ref{A_z_semiinf_minus_infty_to_0})
together with Eq.~(\ref{A_x_semiinf_minus_infty_to_0}), or Eq.~(\ref{A_z_semiinf_0_to_infty}) together with
Eq.~(\ref{A_x_semiinf_0_to_infty}), excluding the paraxial region around $\theta=0$. This exclusion is meaningless
from the point of view of the calculation, because as the distance from the center of coordinates $r$ increases
this region reduces to a singular point. The spherical potential vectors for the RT and LT configurations differ
only by sign, so both yield the same result for $P_{rad}^+$. Using the RT configuration, we rewrite
\begin{equation}
A_{z}=\mu_0 F_{(z)}(\theta,\varphi) G(r),
\label{A_z_sphere_semiinf}
\end{equation}
where
\begin{equation}
F_{(z)}(\theta,\varphi)=-I^+ d \cos\varphi\frac{\sin\theta}{1-\cos\theta},
\label{F_12_semiinf}
\end{equation}
is the directivity function.
We calculate now the radiating electric and magnetic fields, i.e. the part of the fields which decays like $1/r$,
by approximating $\boldsymbol{\nabla}\simeq -jk\mathbf{\widehat{r}}$, obtaining
\begin{equation}
\mathbf{H}_{(z)}=jk F_{(z)}(\theta,\varphi)\sin\theta G(r) \boldsymbol{\widehat{\varphi}},
\label{H_12_semiinf}
\end{equation}
and $\mathbf{E}_{(z)}=\eta_0\mathbf{H}_{(z)}\times\mathbf{\widehat{r}}$.
Now we rewrite (\ref{A_x_semiinf_minus_infty_to_0}):
\begin{equation}
A_x=\mu_0 F_{(x)}(\theta,\varphi) G(r),
\label{A_x3_semiinf_1}
\end{equation}
where the directivity function $F_{(x)}$ is
\begin{equation}
F_{(x)}(\theta,\varphi)= -I^+ d,
\label{F_3_semiinf}
\end{equation}
The fields are $\mathbf{H}_{(x)}=(-jk\mathbf{\widehat{r}})\times (A_x\mathbf{\widehat{x}})/\mu_0$
\begin{equation}
\mathbf{H}_{(x)}=
-jk(\cos\theta\cos\varphi\boldsymbol{\widehat{\varphi}} + \sin\varphi\boldsymbol{\widehat{\theta}})G(r)F_{(x)},
\label{H_3_semiinf}
\end{equation}
and $\mathbf{E}_{(x)}=\eta_0\mathbf{H}_{(x)}\times\mathbf{\widehat{r}}$.
Adding up the fields $\mathbf{H}_{(z)}+\mathbf{H}_{(x)}$ we obtain
\begin{equation}
\mathbf{H}^+=-jkG(r)I^+d[\boldsymbol{\widehat{\varphi}}\cos\varphi - \boldsymbol{\widehat{\theta}}\sin\varphi].
\label{H_semiinf}
\end{equation}
We named it $\mathbf{H}^+$, because it is the radiating field of a forward wave.
and
\begin{equation}
\mathbf{E}^+=\eta_0\mathbf{H}^+\times\mathbf{\widehat{r}}
\label{E_semiinf}
\end{equation}
resulting in Poynting vector $\mathbf{E}^+\times\mathbf{H}^{+*}$ :
\begin{equation}
\mathbf{S}^+=30\,\Omega \frac{(kd)^2\mathbf{\widehat{r}}}{4\pi r^2}|I^+|^2,
\label{S_semiinf_1}
\end{equation}
So that the total radiated power for a forward wave is
\begin{equation}
P_{rad}^+=\int_0^{2\pi}\int_0^{\pi}\sin\theta d\theta d\varphi r^2 \mathbf{\widehat{r}}\cdot\mathbf{S}=
30\,\Omega |I^+|^2 (kd)^2 
\label{P_rad_semiinf}
\end{equation}
It is clear from Eqs.~(\ref{H_semiinf}) and (\ref{S_semiinf_1}) that this is an isotropic radiation. Calculating
$D^+=4\pi r^2 S_r^+/P_{rad}^+$, one obtains
\begin{equation}
D^+=1,
\label{D_plus_semiinf}
\end{equation}
so that we encounter again an isotropic radiation, but contrary to the case shown in Section~\ref{matzner_comparison},
here the polarization is linear. This is possible because the radiation field is not the only field far from the
origin, and the near plane wave is also present, see Figures~\ref{semi_infinite_RT} and \ref{semi_infinite_LT}.

\begin{figure}[!tbh]
\includegraphics[width=9cm]{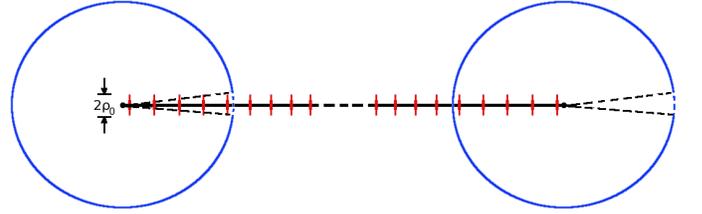}
\caption{Very long TL, carrying a forward wave, for which one considers each termination as the termination of
a semi-infinite TL. Each termination radiates the power $30\,\Omega |I^+|^2 (kd)^2$, so that the whole TL radiates
$60\,\Omega |I^+|^2 (kd)^2$, the asymptotic value in Eq.~(\ref{P_rad}).}
\label{RT_LT}
\end{figure}

\begin{figure}[!tbh]
\includegraphics[width=9cm]{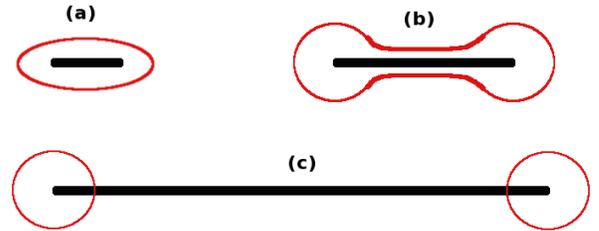}
\caption{A schematic diagram showing the connection between the radiation of finite TL and semi-infinite TL
carrying a forward wave. The red lines are hand drawn and go around the element considered for the
calculation of the radiation, so their shape is meaningless. Panel (a) shows a short TL radiating
the power $60\,\Omega (kd)^2|I^+|^2\left[1-\sinc(4kL)\right]$ according to Eq.~(\ref{P_rad}).
Panel (b) shows a TL longer than several wavelengths, for which $|\sinc(4kL)|\ll 1$, which practically
radiates $60\,\Omega (kd)^2|I^+|^2$, but still considered a single radiating element.
Panel (c) shows a very long TL, like in Figure~\ref{RT_LT}, which is analyzed as two separate radiating
elements (as shown in Figures~\ref{semi_infinite_RT} and \ref{semi_infinite_LT}), each one radiating
$30\,\Omega (kd)^2|I^+|^2$, according to Eq.~(\ref{P_rad_semiinf}), in total the same as in panel (b).}
\label{finite_to_semiinfinite}
\end{figure}

It is worthwhile at this point to understand the connection between the radiation of a finite TL and a
semi-infinite TL. In reality, a semi-infinite TL is a very long TL, for which we analyze the termination
near to ``our'' side, while someone else analyzes the termination near to ``his/her side'', as shown in
Figure~\ref{RT_LT}.
Figure~\ref{finite_to_semiinfinite} shows schematically how the power radiated by a TL carrying a forward wave,
gradually changes as the TL length increases.

Till here we considered only a forward wave, and that is what one usually considers for a semi-infinite TL from
$z=0$ to $\infty$ (LT case), but for the RT
case terminated by a non matched load, one can have both forward and backward waves. The generalization for this case
is done like in Section~II-B, and the total radiated power in presence of a forward and backward wave is
\begin{equation}
P_{rad}=30\,\Omega (kd)^2 (|I^+|^2+|I^-|^2),
\label{P_rad_semiinf_f_b}
\end{equation}
so that the interference between the waves does not contribute to the radiated power, similarly
to the case of a finite TL.

\section{Radiation resistance}
\label{model}

The radiation resistance has already been worked out in Section~\ref{storer_comparison}, for comparison with
\cite{storer}. It is defined by
$r_{\text{rad}}=P_{\text{rad}}/|I|^2$, $P_{\text{rad}}$ being the total power radiated by the TL,
and $I$ the current at the generator side (in Figure~\ref{loaded_TL} it is $I(-L)$). The radiation resistance
is given in Eq.~(\ref{r_rad_any_waves}) and is identical to Eq.~(5) in \cite{storer}.

\begin{figure}[!tbh]
\includegraphics[width=9cm]{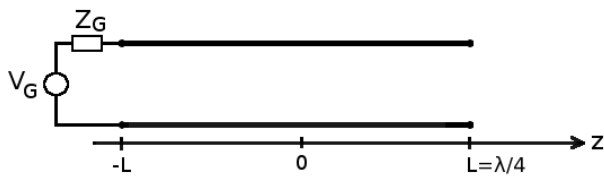}
\caption{Open ended transmission line, of length $2L=\lambda/2$. The current is 0 at both TL ends. The radiation
resistance in Eq.~(\ref{r_rad_any_waves}) (or Eq.~(5) in \cite{storer}) fails in this case, resulting infinity.}
\label{resonant_TL}
\end{figure}

However, we remark that $r_{\text{rad}}$ in Eq.~(\ref{r_rad_any_waves}) goes to infinity if $|\Gamma|=1$ and
$4kL-\angle\Gamma=2\pi n$ ($n$ integer). For example if $Z_L$ in Figure~\ref{loaded_TL} is $\infty$ (open TL),
$\Gamma=1$, so that $I(L)=0$. If the length of the TL $2L$ is a multiple integer of $\lambda/2$,
also the current at the generator side $I(-L)=0$. This is shown schematically in Figure~\ref{resonant_TL},
for $n=1$.
This case represents resonance (infinite VSWR) and current at generator side
0. This of course does not mean that the generator does not have to compensate for the radiated power, but
rather a fail of the small radiation losses approximation. In such case $P^+=|I^+|^2Z_0$ equals $P^-=|I^-|^2Z_0$
so that the net power carried by the TL $P=P^+ - P^-=0$, hence the radiated power $P_{rad}$ is infinitely
bigger than the net power $P$ transferred by the TL.

We derive here a more robust radiation resistance, valid for any TL configuration. We still assume 
$P^+_{rad}\ll P^+$ (small relative losses), but the total radiated power $P_{rad}$ is allowed to be bigger than the
net power $P$. Given the fact that the interference between the forward and backward waves does not contribute
to the radiated power (see (Eq.~\ref{P_rad1_forw_back})), we may consider the separate loss of the forward
or backward wave.

The relation $P_{rad}^+/P^+$ in Eq.~(\ref{rel_rad_calc}), equal also to $P_{rad}^-/P^-$, is written as if $P^+$ would be
a constant, but $P^+$ is only approximately constant for $P^+_{rad}\ll P^+$. Looking at the configuration
in Figure~\ref{loaded_TL}, by conservation of energy, the power radiated by a forward wave $P^+_{rad}$
must be the difference $P^+(-L)-P^+(L)$, which is small relative to the individual values of $P^+(-L)$
and $P^+(L)$. We may therefore express $P^+(L)=P^+(-L)-P^+_{rad}$, but considering $P^+_{rad}\ll P^+$, this
may be written as $P^+(L)= P^+(-L)\left[1-P^+_{rad}/P^+\right]$, or more conveniently
$P^+(L)= P^+(-L)\exp(-P^+_{rad}/P^+)$.

The small decay factor $P^+_{rad}/P^+$ (or
$P^-_{rad}/P^-$ for the backward wave) is minus twice the imaginary part of the TL electrical length
$2 \text{Im}\{\Theta\}$ according to Eq.~(\ref{rel_rad_sim}), so we may describe the dynamics of $P^+$ or $P^-$ along the TL
\begin{align}
P^+(L)=P^+(-L) e^{2 \text{Im}\{\Theta\}}
\label{P_plus_change}
\end{align}
\begin{align}
P^-(-L)=P^-(L) e^{2 \text{Im}\{\Theta\}}
\label{P_minus_change}
\end{align}
where Im$\{\Theta\}<0$ always. Given $P^{\pm}$ are proportional to $|I^{\pm}|^2$ respectively, the forward
and backward currents decay according to $\text{Im}\{\Theta\}$, in addition to their accumulated phase, so we
express
\begin{equation}
I^+(L)=I^+(-L)e^{-j2kL} e^{\text{Im}\{\Theta\}},
\label{I_plus_change}
\end{equation}
\begin{equation}
I^-(-L)=I^-(L)e^{-j2kL} e^{\text{Im}\{\Theta\}},
\label{I_minus_change}
\end{equation}
At the load side $I^-(L)=-\Gamma I^+(L)$, so using Eqs.~(\ref{I_plus_change}) and (\ref{I_minus_change}), we express
the total current near the generator $I(-L)=I^+(-L)+I^-(-L)$:
\begin{align}
I(-L)=I^+(-L)\left[1-\Gamma e^{-j4kL} e^{2 \text{Im}\{\Theta\}}\right]
\label{I_minus_L_total}
\end{align}
which for $|\text{Im}\{\Theta\}|\ll 1$ can be written:
\begin{align}
I(-L)=I^+(-L)\left[1-\Gamma e^{-j4kL} (1+2 \text{Im}\{\Theta\})\right]
\label{I_minus_L_total_1}
\end{align}
Using Eqs.~(\ref{r_rad_def}), (\ref{P_rad1_forw_back}) and the relation $-2 \text{Im}\{\Theta\}=P^+_{rad}/P^+$
from Eq.~(\ref{rel_rad_calc}) we obtain a more accurate, explicit expression for the radiation resistance
\begin{align}
r_{\text{rad}}=\frac{60\,\Omega (kd)^2 \left[1-\sinc(4kL)\right](1+|\Gamma|^2)}{|1-\Gamma e^{-4jkL} \{1- (60\,\Omega/Z_0)(kd)^2\left[1-\sinc(4kL)\right] \}|^2}.
\label{r_rad_any_waves_acc}
\end{align}
If $\Gamma e^{-4jkL}$ is far from 1, the last term in the denominator is negligible, and one recovers the
approximate Eq.~(\ref{r_rad_any_waves}). On the other hand if $\Gamma e^{-4jkL}=1$ (resonance), one obtains
\begin{align}
r_{\text{rad}}=\frac{2Z_0^2}{60\,\Omega (kd)^2\left[1-\sinc(4kL)\right] }.
\label{r_rad_any_waves_acc_res}
\end{align}
which is big, because $kd\ll 1$, but not infinite. For the special case described in Figure~\ref{resonant_TL},
$\Gamma=1$ and $4kL=2\pi$, the $\sinc$ function is 0, so that $r_{\text{rad}}$ reduces to
$2Z_0^2/[60\,\Omega\, (kd)^2]$.

\section{Conclusions}

We derived in this work a general radiation losses model for two-conductors transmission lines
(TL) in free space. We considered any combination of forward and backward waves (i.e. any
termination), and also any TL length, analyzing infinite, semi-infinite and finite TL.

One important finding is that the interference between forward and backward waves does not
contribute to the radiated power (Eq.~(\ref{P_rad1_forw_back})), which has been also validated
by the comparisons with \cite{Manneback,storer,Bingeman_2001,Matzner,Guertler}, in the sense that those comparisons
would have failed if Eq.~(\ref{P_rad1_forw_back}) were incorrect.

This property
allowed us to consider the separate losses for the forward an backward wave for the calculation
of the radiation resistance in Section~\ref{model}. This radiation resistance reduces correctly
far from resonance to this calculated in \cite{storer} (Eq.~\ref{r_rad_any_waves}), but handles correctly
the resonant case.

Another novelty of this work is the analysis of the semi-infinite TL which clearly shows
that the radiation from TL is mainly a termination effect. We found an isotropic radiation
from the semi-infinite TL, which is possible due to the fact that the radiation fields
are not the only far fields, as shown in Figures~\ref{semi_infinite_RT} and \ref{semi_infinite_LT}.
The semi-infinite TL radiation results are consistent with finite TL results, so that 
a very long TL can be regarded as two semi-infinite TLs, as shown in
Figure~\ref{finite_to_semiinfinite}.

Although previous works \cite{storer,Manneback,Bingeman_2001,Nakamura_2006,carson}
considered exclusively the twin lead cross section, the formalism developed in this work is valid
for any cross section. We showed this in Section~\ref{ANSYS} by successfully comparing the analytic results
with simulation of ANSYS-HFSS commercial software for a parallel cylinders cross section
(Figure~\ref{parallel_cylinders}), in which the radius was not small relative to the distance
between the centers of the cylinders. Appendix~B explains how to calculate the parameters
needed to derive the radiation for any TL cross section.

Some comments on the generalization of this research for TL in dielectric insulator.
The case of TL in dielectric insulator is solvable analytically, but much more involved than
the free space case. The fact that the TL propagation wavenumber $\beta$ is different form
the free space wavenumber $k$ by itself complicates the mathematics, but in addition it comes
out that one needs to consider in this case also polarization currents, which further complicate
the results. Radiation from TL in dielectric insulator will be published separately, as Part~II of this
study.


%



\appendices
\renewcommand\thefigure{\thesection.\arabic{figure}}
\setcounter{figure}{0}
\renewcommand\theequation{\thesection.\arabic{equation}}
\setcounter{equation}{0}
\section{Far vector potential of separated two-conductors transmission line}

We show in this appendix that for the purpose of calculating the far fields from a two ideal conductor
transmission line (TL) in free space, having a well defined separation between the conductors,
as shown in Figure~\ref{config}, one can use an equivalent twin lead, provided the separation is much
smaller than the wavelength.

For simplicity we use a forward wave (propagating like $e^{-jkz}$), but the same conclusion is valid
for a combination of waves. In the far field the $z$ directed magnetic potential vector $A_z$ is expressed as
\begin{equation}
A_z=\mu_0\int_{z_1}^{z_2} dz' \oint dc\, K_z(c)e^{-jkz'}G(R)
\label{A_z_basic}
\end{equation}
where the $dz'$ integral goes on the whole length of the TL,
\begin{equation}
G(s)=\frac{e^{-jks}}{4\pi s}
\label{Green}
\end{equation}
is the 3D Green's function, $K_z$ is the surface current distribution as function of the
contour parameter $c$ (i.e. $c_1$ and $c_2$, see Figure~\ref{config}) which is known from electrostatic considerations, and $R$ is
the distance from the integration point on the contour of the conductors to the observer:
\begin{equation}
R=\sqrt{(x-x'(c))^2+(y-y'(c))^2+(z-z')^2}.
\label{R_basic}
\end{equation}
Changing variable $z''=z'-z$ in Eq.~(\ref{A_z_basic}), one obtains
\begin{equation}
A_z=\mu_0 e^{-jkz}\int_{z_1-z}^{z_2-z} dz'' \oint dc\, K_z(c)e^{-jkz''}G(R),
\label{A_z_basic_01}
\end{equation}
redefining $R=\sqrt{(x-x'(c))^2+(y-y'(c))^2+(z'')^2}$.
For a far observer, at distance $\rho\equiv\sqrt{x^2+y^2}$ from the TL, so that $\rho$ is much bigger than the
transverse dimensions of the TL one approximates $R$ in cylindrical coordinates as
\begin{equation}
R\simeq r- \frac{\rho}{r} \left[x'(c)\cos\varphi+y'(c)\sin\varphi\right],
\label{R_basic_1}
\end{equation}
where $r(z'')\equiv\sqrt{(z'')^2+\rho^2}$. We keep for now everything in cylindrical coordinates, to be able
to handle infinite or semi-infinite lines. Using this in Eq.~(\ref{A_z_basic_01}), one obtains
\begin{align}
A_z=&\mu_0 e^{-jkz}\int_{z_1-z}^{z_2-z} dz''\frac{e^{-jk[z''+r(z'')]}}{4\pi r(z'')} \notag \\
    & \oint dc\, K_z(c)e^{jk(\rho/r)[x'(c)\cos\varphi+y'(c)\sin\varphi]}.
\label{A_z_basic_1}
\end{align}
We consider the higher modes to be in deep cutoff, so that $kx'(c),ky'(c)\ll 1$, hence
\begin{align}
A_z\approx&\mu_0 e^{-jkz}\int_{z_1-z}^{z_2-z} dz''\frac{e^{-jk[z''+r(z'')]}}{4\pi r(z'')} \oint dc\, K_z(c) \notag \\
          &\left\{1+jk(\rho/r)[x'(c)\cos\varphi+y'(c)\sin\varphi]\right\}.
\label{A_z_basic_2}
\end{align}
Separating the contour integral $\oint dc=\oint dc_1+\oint dc_2$, where $c_{1,2}$ are the contours of the ``upper''
and ``lower'' conductors respectively (see Figure~\ref{config}), and using
\begin{equation}
\oint dc_1 K_z(c_1)=-\oint dc_2 K_z(c_2)=I^+
\label{I^+}
\end{equation}
so that the integral on each surface current distribution results in the total current, which we call $I^+$, because
it represents a forward wave. Given that for a two-conductors TL there is only one (differential) TEM mode, this current is equal, but
with opposite signs on the conductors. We may define the 2D vector $\boldsymbol{\rho}(c)\equiv(x'(c),y'(c))$,
from which one defines the vector distance between the center of the surface current distributions
\begin{equation}
\mathbf{d}\equiv \left[\oint dc_1 K_z(c_1)\boldsymbol{\rho}(c_1)+\oint dc_2 K_z(c_2)\boldsymbol{\rho}(c_2)\right]/I^+.
\label{vec_d}
\end{equation}
From this point, the original cross section is relevant only for calculating the equivalent separation vector
$\mathbf{d}$ in the twin lead representation, and the remaining calculation bases solely on this {\it twin lead}
representation. In appendix~B we show examples for the calculation of the twin lead equivalent for given
cross sections.

Using the twin lead representation, Eq.~(\ref{A_z_basic_2}) may be rewritten
\begin{align}
A_z=&\mu_0 e^{-jkz}I^+ jk[d_x\cos\varphi+d_y\sin\varphi] \notag \\
    &\int_{z_1-z}^{z_2-z} dz''\frac{e^{-jk[z''+r(z'')]}}{4\pi r(z'')} \frac{\rho}{r(z'')} ,
\label{A_z_basic_3}
\end{align}
where $d_x$ and the $d_y$ are the $x$ and $y$ components of the vector $\mathbf{d}$. This represents
a twin lead, as shown in Figure~\ref{conf_tl_rad}, and is actually a 2D dipole approximation of the TL. 
Without loss of generality, one redefines the $x$ axis to be aligned with $\mathbf{d}$, so that $d_x=d$ and $d_y=0$,
obtaining
\begin{equation}
A_z=\mu_0 e^{-jkz}I^+ jkd\cos\varphi\int_{z_1-z}^{z_2-z} dz''\frac{e^{-jk[z''+r(z'')]}}{4\pi r} \frac{\rho}{r(z'')} ,
\label{A_z_basic_4}
\end{equation}
which is equivalent of having a current $I^+ e^{-jkz}$ confined on the conductor at $x=d/2$ 
and the same current confined on the conductor $x=-d/2$
but defined in the opposite direction, representing a twin lead (see Figure~\ref{conf_tl_rad}).
We are interested in radiation, so we require the observer to be many wavelengths far from the TL:
$k\rho\gg 1$ and $kr\gg 1$, so that Eq.~(\ref{A_z_basic_4}) may be further simplified to
\begin{equation}
A_z= -\frac{\mu_0I^+ e^{-jkz}d\cos\varphi}{4\pi} \frac{\partial}{\partial\rho}\int_{z_1-z}^{z_2-z} dz''\frac{e^{-jk[z''+r(z'')]}}{r(z'')}.
\label{A_z_basic_5}
\end{equation}
The $dz''$ integral results in the exponential integral function $\Ei$ as follows
\begin{align}
A_z=&-\frac{\mu_0I^+ e^{-jkz}d\cos\varphi}{4\pi} \notag \\
    & \left. \frac{\partial}{\partial\rho}\Ei\left(-jk\left[z''+\sqrt{(z'')^2+\rho^2}\right]\right)\right|_{z_1-z}^{z_2-z},
\label{A_z_basic_6}
\end{align}
where the $\Ei$ function satisfies $d\Ei(s)/ds=e^s/s$.

The twin lead geometry also allows us to use simple models for the
termination currents in the $x$ direction (see Figure~\ref{conf_tl_rad}), defining the $x$ component of the magnetic vector
potential, calculated as:
\begin{equation}
A_{x\, 1,2}=\pm\mu_0I^+\int_{-d/2}^{d/2}dx' e^{-jkz_{1,2}}G(R_{1,2})
\label{A_x12}
\end{equation}
where the indices 1,2 denote the contributions from the termination currents at $z_{1,2}$, respectively,
see (see Figure~\ref{conf_tl_rad}) and the distances $R_{1,2}$ of the far observer from the terminations 
may be expressed in spherical coordinates, as follows:
\begin{equation}
R_{1,2}\simeq r - z_{1,2}\cos\theta - x'\sin\theta\cos\varphi,
\label{R_12}
\end{equation}
The integral (\ref{A_x12}) is carried out for $kd\ll 1$, resulting in
\begin{equation}
A_{x\, 1,2}=\pm\mu_0I^+d G(r)e^{-jkz_{1,2}(1-\cos\theta)}
\label{A_x12_1}
\end{equation}

\renewcommand\thefigure{\thesection.\arabic{figure}}
\setcounter{figure}{0}
\renewcommand\theequation{\thesection.\arabic{equation}}
\setcounter{equation}{0}
\section{Computation of radiation parameters}

\begin{figure}[!tbh]
\includegraphics[width=9cm]{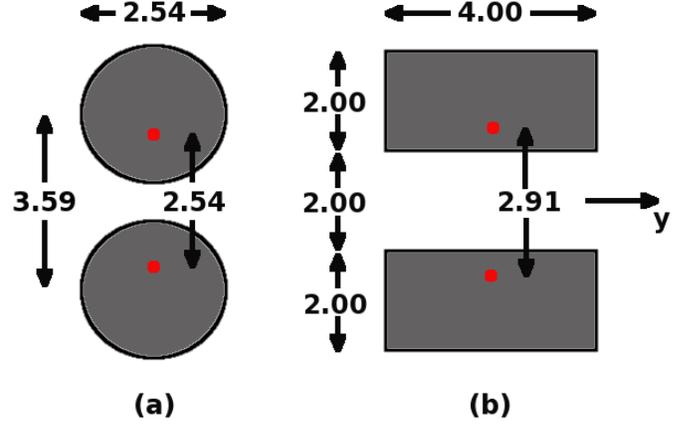}
\caption{Panel (a) shows a cross section of circular shaped conductors and panel (b) shows a cross section
of rectangular shaped conductors. The sizes are in units of cm. The $y$ axis is horizontal, and the $x$ axis
for each cross section is the symmetry axis. The red points show the current images which 
define the twin lead representation, and are referred further on.}
\label{cross_sections}
\end{figure}

\begin{figure}[!tbh]
\includegraphics[width=9cm]{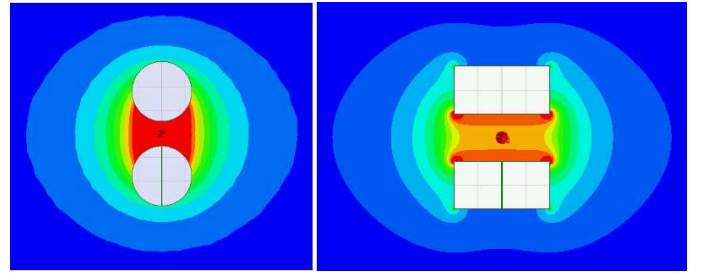}
\caption{The magnitude of the electric field, the hottest color representing high field and coldest
color low (close to 0) field, for the circular shaped conductors cross section in the left panel and
for the rectangular shaped conductors in the right panel. The red scars in the middle of the plots
show the coordinate's origin.}
\label{maxwell_fields}
\end{figure}

To calculate the power radiated from a TL, of any cross section, one needs the separation vector   
$\mathbf{d}$, defined in Figure~\ref{config}, calculated from Eq.~(\ref{vec_d}). If one needs the
normalized radiation due to a forward wave,
one also needs to know the characteristic impedance $Z_0$. Those are obtained with the aid of the
ANSYS 2D ``Maxwell'' simulation, from an electrostatic analysis. We ran the 2D ``Maxwell'' simulation
on two cross sections shown in Figure~\ref{cross_sections}.
It is to be mentioned that we know the analytic solution for the cross section in panel (a) from image
theory \cite{orfanidis}, so that it can be used as a test for the quality of the numerical simulation. The magnitude
of the electric fields measured for those cross sections is shown in Figure~\ref{maxwell_fields}.
We remark that the surface currents, which are proportional to the (tangential) magnetic field
on the conductors are also proportional to the (normal) electric field on the conductors. Given
the surface currents $K_z(c_1)>0$ and $K_z(c_2)<0$ in Eq.~(\ref{vec_d}), and using the field
intensity which is positive, $E(c_1)$ is proportional to $K_z(c_1)$ and $E(c_2)$ is proportional to $-K_z(c_2)$,
we may calculate the separation vector $\mathbf{d}$, using Eqs.~(\ref{I^+}) and (\ref{vec_d}), after replacing
$K_z(c_1)$ by $E(c_1)$ and $K_z(c_2)$ by $-E(c_2)$ as follows
\begin{equation}
\mathbf{d}=\frac{\oint dc_1 E(c_1)\boldsymbol{\rho}(c_1)-\oint dc_2 E(c_2)\boldsymbol{\rho}(c_2)}{\oint dc_1 E(c_1)}.
\label{vec_d_num}
\end{equation}
For the cross sections in Figure~\ref{cross_sections}, the ``positive'' and ``negative'' conductors are symmetric,
so that a given location vector $\boldsymbol{\rho}(c_2)$ on the negative conductor, is the minus of the {\it corresponding}
location vector $\boldsymbol{\rho}(c_1)$ on the positive conductor, and by symmetry the magnitudes of the electric
fields $E(c_1)=E(c_2)$, so that we may drop the second integral in the numerator of Eq.~(\ref{vec_d_num}), and multiply
the result by 2. Also by symmetry the $y$ component of $\mathbf{d}$ comes out 0, so that $d=|\mathbf{d}|=d_x$ represents
the distance between the ``image'' currents in the twin lead model, and we obtained $d=2.54$~cm for the circular cross
section (compares well with Eq.~(\ref{d})), and $d=2.91$~cm for the rectangular cross section, as shown in Figure~\ref{cross_sections}.

We obtained from the 2D ``Maxwell'' simulation also the per unit length capacitances, which came out
31.5~pF/m and 33.51~pF/m for the circular and rectangular cross sections, respectively. The characteristic
impedance $Z_0$ is calculated by $1/(Cc)$, where $c$ is the velocity of light in vacuum and $C$ is the 
per unit length capacitance, and come out 105.8$\,\Omega$ (compares well with Eq.~(\ref{Z0})) and
99.51$\,\Omega$ for the circular and rectangular cross sections, respectively.

The procedure described in this section can be done for any cross section, and it supplies all the values
needed to calculate the normalized radiation in Eq.~(\ref{rel_rad_calc}). Its accuracy can be found by
comparing the values obtained for $d$ and $Z_0$ for the circular cross section, with the theoretical values
obtained from image theory and they fit with an inaccuracy of less than 0.5\%.

\begin{IEEEbiography}[{\includegraphics[width=1in,height=1.25in,clip,keepaspectratio]{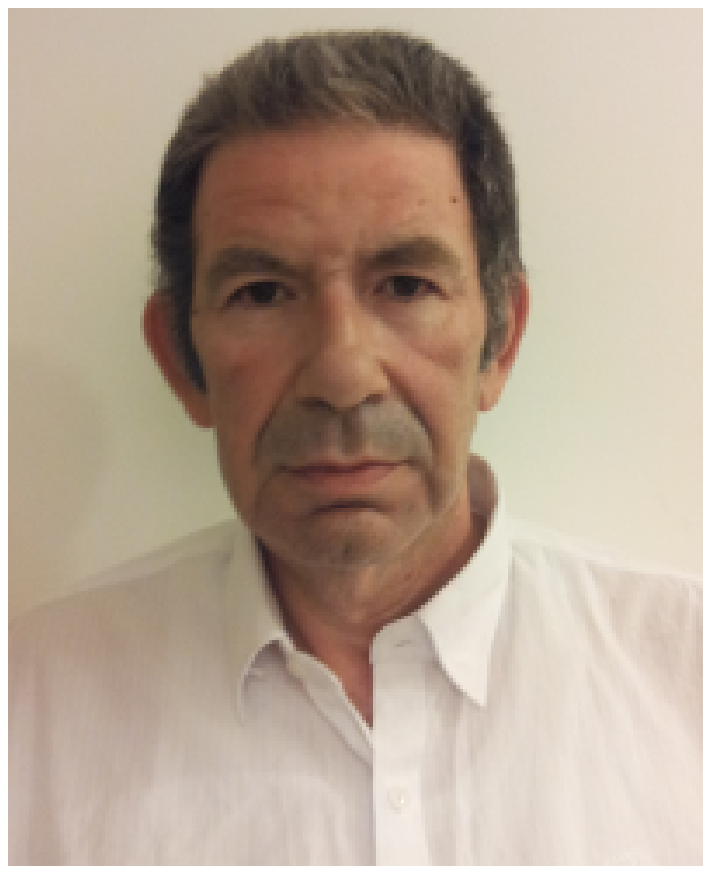}}]{Reuven Ianconescu}
graduated B.Sc (Summa cum laude) in 1987 and PhD in 1994, both in Electrical engineering at Tel Aviv university.
He worked many years in R\&D in the hi-tech industry, did a post-doc in the Weizmann Institute of Science, and is currently an academic staff member in Shenkar college of engineering and design.
His research interests are: guided EM propagation, radiation and dynamics of charges and electrohydrodynamics.
\end{IEEEbiography}

\begin{IEEEbiography}[{\includegraphics[width=1in,height=1.25in,clip,keepaspectratio]{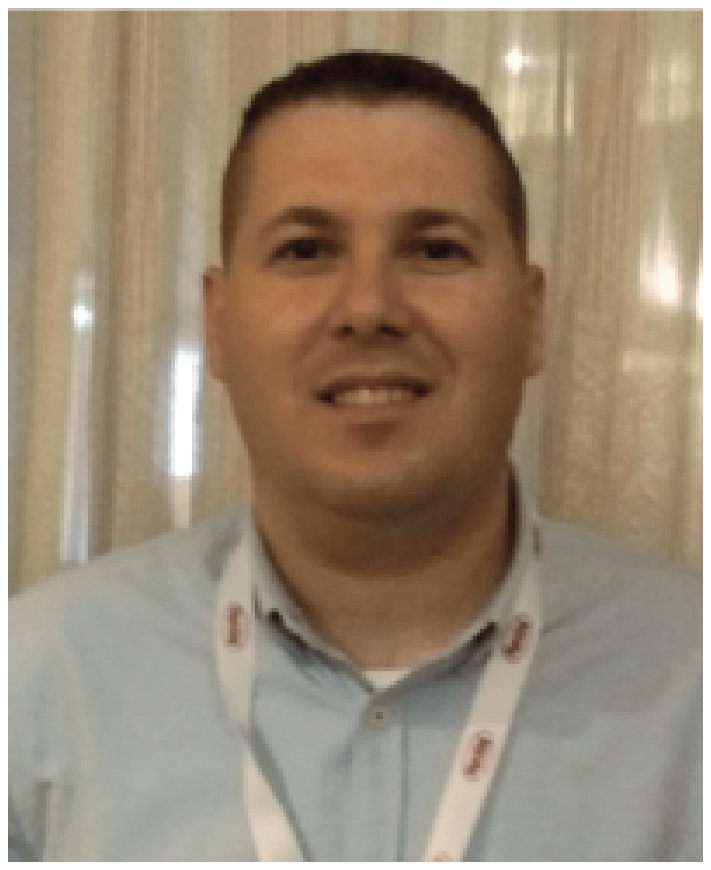}}]{Vladimir Vulfin}
has over 13 years of research experience in the areas of electromagnetic engineering. Since 2004 he worked in different companies as Electromagnetics Engineer. His job duties included design, simulations, implementations and measurements of antennas and microwave devices. Parallel to this, Vladimir was active as an external lecturer and instructor in different universities and colleges in Israel. He lead more than 100 degree final projects for students in Electrical Engineering Departments. Vladimir received the B.Sc and M.Sc. degrees in Electrical Engineering. Currently he is working towards his Ph.D degree at Ben Gurion University in the area of antennas and metamaterials.
\end{IEEEbiography}

\end{document}